\documentclass[11pt]{article}

\usepackage{mathtools}
\usepackage{amssymb}
\usepackage{dsfont}
\usepackage{slashed}
\usepackage{fullpage}
\usepackage[colorlinks=true,linkcolor=blue,citecolor=magenta,linktocpage=true]{hyperref}
\usepackage{titlesec}
\titleformat*{\section}{\normalsize\bfseries}
\titleformat*{\subsection}{\normalsize\bfseries}
\titleformat*{\subsubsection}{\normalsize\bfseries}
\usepackage{cite}
\usepackage{amsmath,amssymb,braket,tikz}
\usetikzlibrary{tikzmark,calc}

\usepackage[english]{babel}
\usepackage{csquotes}
\MakeOuterQuote{"}

\DeclareMathAlphabet{\bbvar}{U}{BOONDOX-ds}{m}{n}

\makeatletter
\renewcommand{\@dotsep}{10000}
\makeatother

\def\be#1\ee{\begin{align}#1\end{align}}
\def\nn{\nonumber}

\def\f{\frac}
\def\eps{\varepsilon}

\def\C{\mathcal{C}}

\def\G{\mathcal{G}}
\def\H{\mathcal{H}}

\def\L{\mathcal{L}}
\def\N{\mathcal{N}}
\def\O{\mathcal{O}}

\def\R{\mathcal{R}}

\def\M{\mathcal{M}}

\def\tt{\tilde{t}}
\def\heq{\,\hat{=}\,}
\def\nb{n}

\newcommand{\cO}{{\mathcal O}}
\newcommand{\cS}{{\mathcal S}}
\newcommand{\cL}{{\mathcal L}}

\renewcommand{\sl}{{\mathfrak{sl}}}
\newcommand{\cw}{{\mathfrak{Witt}}}

\def\rd{\textrm{d}}
\def\pp{\partial}

\def\beq{\begin{eqnarray}}
\def\eeq{\end{eqnarray}}
\def\be{\begin{equation}}
\def\ee{\end{equation}}
\def\vphi{\varphi}
\newcommand{\sh}{{\mathfrak{sh}}}

\newcommand{\so}{{\mathfrak{so}}}

\numberwithin{equation}{section}

\begin{document}

\title{\Large{\textbf{\sffamily Schr\"{o}dinger Symmetry in Gravitational Mini-Superspaces}}}
\author{\sffamily Jibril Ben Achour\;$^{1,2}$, Etera R. Livine\;$^{2}$, Daniele Oriti\;$^{1}$ and Goffredo Piani\;$^{ 1,3}$}
\date{\small{\textit{$^{1}$ Arnold Sommerfeld Center for Theoretical Physics, Munich, Germany, \\
$^{2}$ Univ de Lyon, ENS de Lyon, Laboratoire de Physique, CNRS UMR 5672, Lyon 69007, France,\\
 $^{3}$ Dipartimento di Fisica e Astronomia, Università di Bologna, Bologna, Italy}}}

\maketitle

\begin{abstract}

We prove that the simplest gravitational
symmetry reduced models
describing cosmology and black holes mechanics are invariant under the Schr\"{o}dinger group.
We consider the flat FRW cosmology filled with a massless scalar field and the Schwarzschild black hole mechanics, construct their  conserved charges using the Eisenhart-Duval (ED) lift method and show that they form a Schr\"{o}dinger algebra.
Our method illustrates how the ED lift and the more standard approach analyzing the geometry of the field space are complementary in revealing different set of symmetries of these systems.
We further identify an infinite-dimensional symmetry for those two models, generated by conserved charges organized in two copies of a $\cw$ algebra.
%
These extended charge algebras provide a new algebraic characterization of 
these homogeneous gravitational sectors. They guide the path to their quantization and open the road to non-linear extensions of quantum cosmology and quantum black holes models in terms of hydrodynamic equations in field space.

\end{abstract}

\thispagestyle{empty}
\newpage
\setcounter{page}{1}

\hrule
\tableofcontents
\vspace{0.7cm}
\hrule

\newpage

\section{Introduction}

Uncovering symmetries of general relativity and gravitational systems beyond the gauge symmetry under space-time diffeomorphisms is a crucial endeavour to understand the theory, both at the classical and quantum levels. This exercise has a clear intrinsic interest, but  also opens the road to build fruitful dictionaries with other physical systems sharing the same symmetries, at least in some given regime. In particular, addressing this question in the context of reduced gravitational models
reveals new key structures for their quantizations and the construction of analogue models of quantum gravity. 

Reduced gravitational models - or mini-superspaces\footnotemark{}, in the jargon of general relativity - are obtained by imposing homogeneity conditions to the gravitational (and matter) fields.
\footnotetext{
Although the terminology of superspace is well-rooted in the study of gravitational models and was introduced to describe the configuration space and phase space of general relativity, we acknowledge that it potentially clashes with the notion of supersymmetry manifolds or supermanifolds, which are sometimes also nicknamed superspaces.
In the context general relativity, the prefix ``super-'' in superspace refers to the structures of the space of metrics over a manifold, which is the structure on a level above considering a space-time with a specific metric. It allows to study the dynamics of general relativity beyond the kinematics of matter fields evolving in one given space-time metric.
To avoid any confusion, the present work does not deal with supergravity or supersymmetric theories, but focuses on mini-superspace models, that is gravitational models for cosmology and astrophysics  obtained by reducing the space of metrics to a finite dimensional space of parameters describing the geometry of space-time.
}
Such restriction focuses on the dynamics of the zero modes of the space-time geometry. The dynamics of the corresponding gravitational field then reduces to a finite-dimensional mechanical system. While such symmetry reduction can appear drastic, it nevertheless provides relevant models for both cosmology and astrophysics. Moreover, this reduction recast the homogeneous gravitational system into a point particle moving in an auxiliary space that is, the space of field configurations (or field space or minisuperspace).
Within this context, one can identify new symmetries of the gravitational zero mode's dynamics which fully encode the evolution of the homogeneous geometry \cite{Pioline:2002qz, Christodoulakis:2013xha, BenAchour:2019ufa, BenAchour:2020njq, Achour:2021lqq, Geiller:2020xze, Achour:2021dtj}.
The exploration of these symmetries was initiated in \cite{BenAchour:2019ufa, BenAchour:2020njq, Achour:2021lqq} for isotropic cosmological models and extended to black hole models in \cite{Geiller:2020xze, Achour:2021dtj}. 
An interesting feature of these symmetries is that they are not realized as standard space-time diffeomorphisms. For instance, the SL$(2,\mathbb{R})$ symmetry discussed in \cite{Achour:2021lqq} amounts at a M\"{o}bius transformation of the time (or radial) coordinate while the components of the metric transform as primary fields with different conformal weights, leading to an anisotropic Weyl rescaling of the metric. Physically, these symmetries map solutions of the homogeneous Einstein's equations onto other gauge-inequivalent ones, as expected from a non-trivial (non-gauge) symmetry.
Understanding the origin of these non standard symmetries in gravitational mini-superspaces is one key goal of this work.



Interestingly,
one can address this question in a fairly elegant manner thanks to the geometrical techniques developed for mechanical systems. The general idea amounts to geometrizing the dynamics of the system under consideration in terms of geodesic motion in the field space.
%
Identifying the conformal isometries of this background provides a geometrization of the symmetries of the system. A first approach consists in considering the space of dynamical fields, whose metric is constructed from the kinetic matrix of the system  \cite{Christodoulakis:2013xha}. This is the approach developed in \cite{Dimakis2015, Terzis:2015mua, Christodoulakis:2018, Christodoulakis:2012eg, Christodoulakis:2013sya} and extended more recently in \cite{Geiller:2022baq} to systematically explore the dynamical symmetries of homogeneous gravity models. This is a very natural approach as it can be motivated also by the fact that the general relativistic dynamics of the gravitational field can be recast in the form of geodesic motion (of a point particle) on field space \cite{Greensite:1995ep}. A second approach makes use of an extended notion of field space, the so-called Eisenhart-Duval (ED) lift \cite{ED}. See \cite{Cariglia:2015bla} for an introduction and \cite{Cariglia:2014dwa} for a detailed presentation. This elegant method has been applied to a large variety of systems, from time-dependent systems, celestial mechanics, inflation and more exotic systems \cite{Duval:1990hj, Gibbons:2010fb, Cariglia:2013efa, Cariglia:2016oft, Cariglia:2018mos, Gibbons:2020nzu, Dhasmana:2021qvw, Fordy:2019vzp, Finn:2020lyw, Sen:2022vig}, and even quantum mechanics \cite{Dunajski:2022ejc}. An interesting fact about this formalism is that the lift metric is equipped with a covariantly constant null Killing vector. It follows that the lift metric can be viewed as a pp-wave, leading to a correspondence between the conformal isometries of pp-waves and dynamical symmetries of mechanical systems \cite{Bekaert:2013fta, Morand:2018tke}. When applied to gravitational mini-superspace models, which already describe homogeneous gravitational fields, this lift can be understood as an elegant second geometrization of gravity. Moreover, this peculiar structure provides a natural host for the geometrical realization of non-relativistic conformal symmetry \cite{Duval:2012qr, Duval:2009vt}. In this work, we shall provide a first application of the ED lift to cosmological and black hole dynamics, revealing in that way the Schr\"{o}dinger symmetry of these gravitational mini-superspaces. Then, comparing the two approaches, one finds that the two formalisms reveal in general different symmetries of the underlying system. We illustrate their complementarity by revealing an additional infinite dimensional symmetry for each system.

Using this geometrization techniques, we show that reduced gravitational models
relevant for cosmology or black holes possess a symmetry under the Schr\"{o}dinger group. This non-relativistic conformal symmetry is a key invariance of classical and quantum mechanics, and also appears in suitable non-linear Schr\"{o}dinger equations describing Bose-Einstein condensates, such as the two dimensional Gross-Pitaevskii equation as well as in many other areas \cite{Nied, Duval:2012qr, Duval:2009vt, Horvathy:2009kz, Son:2008ye, Balasubramanian:2008dm, Taylor:2008tg, Goldberger:2008vg, Duval:2008jg}. The realization of the Schr\"{o}dinger symmetry in these gravitational systems thus provides a new structure useful to organize their quantization. It also stands as new guide to develop non-linear extensions of quantum cosmology and quantum black hole based on symmetries, in close analogy with non-linear Schr\"{o}dinger equations. In particular, this non-linear extension will give hydrodynamic-like equations on field space. We will comment on this point in the discussion of our results.

\medskip

\medskip

This work is organized as follows. In Section~\ref{sec1}, we briefly review the key points entering the construction of gravitational mini-superspace models. Then, we provide a review of the geometrization of the model first in terms of the field space, and then in terms of the ED lift.  
%
In Section~\ref{sec: FRW}, we present the Schr\"{o}dinger symmetry of the flat  FRW cosmology filled with a massless scalar field. We derive the Schr\"{o}dinger observables, the symmetry transformations and algebraically solve the dynamics. Section~\ref{sec:BH} presents the results for the Schwarzschild black hole mechanics. Section~\ref{infsym}, we present the infinite dimensional symmetry in our cosmological and black hole models. Finally, Section~\ref{disc} summarizes  our results and discusses  open perspectives.
%

\section{Symmetries from field space geometry}

\label{sec1}

Gravitational mini-superspaces are reduced models of general relativity restricted to classes of space-time metrics defined in terms of a finite number of degrees of freedom, usually based on the assumption of homogeneity.
They describe specific sectors of the full theory, relevant to cosmology and astrophysics. See \cite{Bojowald:2010qpa} for a pedagogical presentation. They are simple mechanical systems, and we review in this section how their dynamics can be given a further geometric translation, not in terms of spacetime geometry, but in terms of motion on the field space (a.k.a. superspace).
In the case of a free mechanical model, the classical trajectories are identified as null geodesics of the field space and the symmetries of the system are given by the conformal Killing vectors of the (super-)metric of the field space.
In the general case, the non-vanishing potential of the mini-superspace will drive the system away from null geodesics.
Nevertheless, one can use the Eisenhart-Duval lift \cite{Cariglia:2015bla, Cariglia:2014dwa}, embedding the field space into a lifted space with 1+1 extra dimensions, to identify once again classical trajectories as null geodesics and symmetries as conformal Killing vectors of the lifted super-metric.


\subsection{Gravitational mini-superspaces and super-metric}

We consider mini-superspaces of general relativity defined by restricting the space-time metrics to a specific ansatz with a finite number of components allowed to vary in terms of a chosen coordinate while the rest of the metric is held fixed and considered as background data. For instance, a typical choice in cosmology is an ansatz of the type:
\be
\rd s^{2}=\gamma_{\mu\nu}\rd x^{\mu}\rd x^{\nu}
= -N^{2}(t)\rd t^{2}
+ \gamma_{ij}[\chi^{a}(t),  x^{i}]\,\rd x^{i}\rd x^{j}\,,
\ee
with no cross terms in $\rd t\rd x^{i}$, where the time component is the lapse function $N$ and the spatial metric $ \gamma_{ij}$ is defined in terms of fields $\chi^{a}$ allowed to vary with the time coordinate $t$ and with a fixed dependance with respect to the space coordinates $x^{i}$.
More generally, allowing for both gravitational and matter degrees of freedom, and denoting $\chi^a(t)$, with $a$ running from 1 to an integer $d$, the degrees of freedom of the mini-superspace ansatz, the  Einstein-Hilbert action for general relativity plus matter typically reduces to a mechanical system of the form:
\begin{align}
S[N, \chi^a, \dot{\chi}^a ; t] =  \ell_P c
\int \rd t\,\left( \frac{1}{2N} g_{ab}(\chi) \dot{\chi}^a \dot{\chi}^b - {N} V(\chi)\right)
\,,
\end{align}
%
where the dynamical fields depend on the coordinate $t$. The action is for a point particle on minisuperspace and is expressed in terms of an affine parameter on the particle trajectories on minisuperspace. These trajectories can be of any signature, with respect to the supermetric $g_{ab}$, thus the affine parameter has no universal spacetime interpretation. In some cases, in particular when one is considering Friedmann dynamics, it can be identified with the "time coordinate" in which the spacetime metric is expressed, but this does not have to be always the case. We nevertheless use, unless specified otherwise, the convention that $t$ is a time coordinate and that we have integrated over the spatial coordinates $x^{i}$.

The pre-factor $\ell_{P} \times c$ comes from the standard $1/\ell_{P}^{2}\propto 1/\hbar G$ factor in front of the Einstein-Hilbert action, combined with the 3d volume $V_{o}$ coming from the integration over the space coordinates (with appropriate cut-off). Since the Planck length $\ell_{P}$ defines the UV scale and the fiducial volume $V_{o}$ defines the IR scale, the dimensionless factor $c$ plays the role of the ratio between UV and IR scales. 

The kinetic terms comes from the extrinsic curvature (of the constant $t$ hypersurfaces) contribution to the Einstein-Hilbert action plus the kinetic terms of the matter Lagrangians. It is expressed  in terms of the $d$-dimensional field space metric $g_{ab}$, or super-metric in short.
The potential $V(\chi^a)$ reflects both the intrinsic curvature (of the constant $t$ hypersurfaces)  and the self-interaction  of the matter fields.
The mini-superspace action is defined by the super-metric and the potential.

The equations of motion of the mini-superspace descend from the stationarity of the action with respect to lapse variations $\delta N$ and to field variations $\delta \chi^{a}$. These should amount to imposing  the Einstein equations on the chosen space-time metric ansatz plus the field equations for matter evolving on that metric. For an arbitrarily chosen metric ansatz, this is clearly a non-trivial statement. Nevertheless, if the space-time metric ansatz is defined by symmetry reduction, i.e. requiring the considered metric to be invariant under a symmetry of the full theory, for instance by imposing a certain number of Killing vectors (thus descending from the diffeomorphism invariance of the Einstein-Hilbert action), then it is automatic that the solutions of the equations of motion of the reduced action are also solutions of the equations of motion of the full theory. Here the equation of motion $\delta S/\delta N=0$ gives the Hamiltonian constraint of the ADM formalism, giving the time evolution of the spatial metric, while the equations of motion $\delta S/\delta \chi^{a}=0$ should lead to the momentum constraints, giving the  projection of the Einstein equations on the spatial slices.
Although we will not attempt a full classification of gravitational mini-superspaces here, we will make sure case by case that the classical solutions of the mini-superspaces studied satisfy indeed the full Einstein equations.

In particular, these models describe straightforward parametrized mechanical systems, invariant under time reparametrizations
\be
\left|\begin{array}{lcrcl}
t &\mapsto& \tt&=&f(t)  \,, \\
N(t) &\mapsto& \tilde{N}(\tt)&=&\f1{\dot{f}(t)}N(t)\,, \\
\chi^{a}(t) &\mapsto& \tilde{\chi}^{a}(\tt)&=&\chi^{a}(t)\,. \\
\end{array}
\right.
\ee
This means that the Hamiltonian of the system vanishes. Indeed, the Legendre transform defines the canonical momenta $\pi_{a}$ and Hamiltonian $H$:
\be
\pi_{a}=\f1N g_{ab}\dot{\chi}^{b}\,,\qquad
H=\pi_{a}\dot{\chi}^{a}-L
=Nh
\quad\textrm{with}\quad
h=\f12g^{ab}\pi_{a}\pi_{b}+V
\,,
\ee
where $g^{ab}$ is the inverse super-metric and the $\chi$ dependence of the Hamiltonian comes from both the super-metric and the potential. The equation of motion with respect to lapse variations simply imposes the Hamiltonian constraint $h=0$ which is a crucial feature of such relativistic system. 

\medskip

We focus in this paper on two specific mini-superspace models:
\begin{itemize}
\item {\bf flat  FRW cosmology filled with a scalar field}:
This model contains two dynamical fields $\chi^a = \{ a, \varphi\}$. The former plays the role of the scale factor in the metric
\be
\rd s^2 = - N^2(t) \rd t^2 + a^2(t) \delta_{ab} \rd x^a \rd x^b
\ee
while the second dynamical field corresponds to an homogeneous scalar field coupled to this metric.

\item  {\bf Schwarzschild black hole mechanics}: The geometry contains two dynamical fields denoted $\chi^a = \{ A, B \}$ and the line element reads
\be
\rd s^2 = \epsilon \left( - N^2(t) \rd t^2 + A^2(t) \rd y^2\right)  + \ell^2_s B^2(t) \left(  \rd \theta^2 + \sin^2{\theta} \rd \phi^2 \right)
\ee
where $\ell_s$ is a fiducial length scale introduced to work with a dimensionless field $B$.
This geometry describes both the interior and the exterior of the black hole. When $\epsilon = +1$, the coordinate $t$ is time-like and labels space-like hypersurfaces foliating the black hole interior, while for $\epsilon = -1$, it is space-like and it labels the time-like foliation of the exterior region, thus playing the role of the radius.

\end{itemize}

On top of being physically relevant gravitational models, in cosmology and astrophysics, these systems have rather generic  mathematical properties.
The flat  FRW cosmology is a free system with a vanishing potential and a conformally flat super-metric, while
the Schwarzschild mechanics has a two-dimensional field space, thus with a conformally flat super-metric, but with a non-vanishing potential.

\medskip

%
For free systems with vanishing potential, the mini-superspace action reads:
\begin{align}
S[N, \chi^a, \dot{\chi}^a ; t] =  \ell_P c
\int \rd t\,\bigg{[} \frac{1}{2N} g_{ab}(\chi) \dot{\chi}^a \dot{\chi}^b \bigg{]}
\,.
\end{align}
This is simply a geodesic Lagrangian: the equations of motion impose that the space-time metric components $\chi^{a}$ follow a null geodesic in field space provided with the super-metric $g_{ab}$. The fact that we must consider null geodesics is imposed by the equation of motion with respect to lapse variations.
Symmetries of the theory will then map the set of null geodesics in field space onto itself.

\medskip

For general mini-superspaces with non-vanishing potential, one can absorb the potential into a field-dependent redefinition of the lapse by defining a proper time coordinate:
\be
\rd \eta\equiv\,NV(\chi)\,\rd t\,.
\ee
%
This choice of lapse leads to the gauge-fixed action:
\begin{align}
S_{gf}[ \chi_a, \dot{\chi}_a ; \eta] = c \ell_P
\int \rd  \eta  \;
\left[ \frac{1}{2}G_{ab}(\chi) \rd_{\eta}\chi^a \rd_{\eta}{\chi}^b  - 1\right]
\,,\qquad\textrm{with}\quad
G_{ab}=V(\chi)g_{ab}\,.
\end{align}
We have absorbed the potential into a conformal rescaling of the super-metric. The constant shift $-1$ in the Lagrangian should usually be dropped out since it would not contribute to the (bulk) equations of motion. Here, it plays the non-trivial role of shifting the physical value of the Hamiltonian.
Indeed, putting aside that constant shift, we recognize once again a geodesic Lagrangian: classical solutions $\chi^{a}(\eta)$ now follows geodesics of the conformally-rescaled super-metric $G_{ab}$. Nevertheless, even though we gauge-fixed the lapse function, one must not forget the original equation of motion with respect to lapse variations. It is indeed ``hidden'' in the definition of the $\eta$ coordinate and it implies the constraint:
\be
\frac{1}{2}G_{ab}(\chi) \rd_{\eta}\chi^a \rd_{\eta}{\chi}^b  - 1 =0\,,
\ee
i.e. it restricts to massive geodesics with fixed mass.
%
In that case, we anticipate the introduction of correction terms proportional to the ``super-mass'' that will have to be added to the conserved charges for null geodesics. 

Another avenue to treat such general mini-superspaces with non vanishing potential is to use the Eisenhart-Duval lift. Indeed, as we shall see in the following, we can embed such systems with potential in a $(d+2)$-dimensional extended field space in order to map them once again on null geodesic Lagrangians. This is will be reviewed in section \ref{sec13}.

\subsection{Symmetries from field space Killing vectors}

\label{sec12}


Let us first study (gravitational) mechanical models with vanishing potential. As we saw above, these free systems are geodesic Lagrangians, whose classical trajectories are null geodesics of the super-metric $g_{ab}$. Since conformal isometries of the super-metric map the set of null geodesics onto itself, and thus define automorphisms of the set of solutions to the equations of motion, conformal Killing vectors (CKVs) of the super-metric naturally define symmetries of the mini-superspace.

More precisely, the action of a free mini-superspace read in Lagrangian form:
\begin{align}
\label{freeac}
S[N, \chi^{a}, \dot{\chi}^{a} ; t] = c \ell_P
\int \rd t\, \frac{1}{2N} g_{ab}(\chi) \dot{\chi}^a \dot{\chi}^b 
\,,
\end{align}
and in Hamiltonian form, after Legendre transform:
\begin{align}
S[N, \chi^{a}, \pi_a ; t] =
c \ell_P
\int \rd t\,\Big{(}\pi_{a}\dot{\chi}^{a} -Nh[\chi,\pi]\Big{)}
\,,
\qquad\textrm{with}\quad
h[\chi,\pi]=\f12g^{ab}\pi_{a}\pi_{b}\,.
\end{align}
Considering a conformal Killing vector $\xi^a \partial_a$ on the field space, with the $\xi^{a}$'s functions of the $\chi^{a}$'s, satisfying
\be
\L_{\xi} g_{ab} = \varphi g_{ab} 
\ee
with the conformal rescaling factor $\varphi (\chi)$, it is a standard result that the scalar product between the velocity vector and the Killing vector is conserved along null geodesics: it is thus a constant of motion. Here, introducing  $\O_{\xi} = \xi^a \pi_a$, a straightforward calculation allows to check that
\be
\{ \O_{\xi}, H \} = - \frac{1}{2} \pi^a \pi^b \L_{\xi} g_{ab} = -  \varphi H
\,\underset{H=0} \heq 0\,.
 \ee
This shows that $\O_{\xi}$ is a weak Dirac observable of the system, it is conserved along classical trajectories and, as a conserved charge, it generates a symmetry of the mini-superspace by Noether's theorem.
Listing all the CKVs of the field space metric  therefore provides a simple way to identify  conserved charges  and symmetries of the mini-superspace model. 
 
However, this procedure only reveals  a subset of the symmetries of the system.
First, the observables constructed above, as $\O_{\xi}$ with  a (conformal) Killing vector $\xi$, are always linear in the momenta $\pi_{a}$. The method clearly misses conserved charges of higher order in the momenta. This could probably be remedied by looking at (conformal) Killing tensors of the super-metric.
Second, this approach can not produce time-dependent conserved charges, or in other words evolving constants of motion, which satisfy:
\be
\label{evolv}
\frac{\rd \O}{\rd \tau} = \partial_{\tau} \O + \{ \O, H \} =0\,,
\ee
because the CKVs $\xi$ depend on the field space variables but not directly on time (and space coordinates).
%
%
To obtain such observables, one has either to adopt the approach followed in \cite{Geiller:2022baq}, or to develop an extension of the field space allowing to treat the affine parameter $t$ on the same footing of the fields $\chi^{a}$. This is actually realized by the Eisenhart-Duval lift which we now review below, in the next section.

\subsection{Eisenhart-Duval lift: back to null geodesics}
\label{sec:EDlift}
\label{sec13}

The Eisenhart-Duval lift is a general method to geometrize any mechanical system, by mapping them onto null geodesic Lagrangians \cite{ED}. The interested reader can refer to \cite{Cariglia:2015bla, Cariglia:2014dwa} for a detailed presentation of the formalism. In the following, we shall review the main points relevant for our investigation of gravitational mini-superspaces.
Let us start with the Lagrangian ansatz:
\be
\label{gen:original}
S[ \chi^a, \dot{\chi}^a ; \tau] =  \ell_p c\int \rd \tau\,
\left( \frac{1}{2} g_{ab}(\chi) \dot{\chi}^a \dot{\chi}^b - {V(\chi)} \right)
\,,
\ee
where we have gauge-fixed the lapse to $N=1$ by introducing the proper time $\rd\tau=N\rd t$. Despite the gauge-fixing, we will keep the Hamiltonian constraint in mind, i.e. that the  value of the Hamiltonian on physical trajectories vanishes, $h=0$.

\medskip

The idea of the  Eisenhart-Duval lift is to introduce another variable $u$ and add a factor $\dot{u}^{2}$ in front of the potential, thereby yielding a geodesic Lagrangian even in presence of non-trivial potentials. More precisely, we introduce a pair of new coordinates $(u,w)$ with a lifted (super-)metric:
\be
\label{gen:EDmetric}
\rd s^{2}_{(ED)}
=2\rd u \rd w -2V(\chi)\rd u^{2}+g_{ab}\rd \chi^{a}\rd\chi^{b}\,,
\ee
leading to a lifted action defined as an integration over a parameter $\lambda$:
\be
\label{gen:lift}
\cS_{(ED)}[\chi(\lambda), u(\lambda),w(\lambda);\lambda]
\,\equiv
\int \rd\lambda\,
\f1{\N}\left(
\dot{u}\dot{w} - {V(\chi)}  \dot{u}^2 + \frac{1}{2} g_{ab}(\chi) \dot{\chi}^a \dot{\chi}^b 
\right)
\,,
\ee
where the dot denotes here the derivative with respect to $\lambda$.
This looks very much like the original Lagrangian and, in fact,  take the same form if $\dot{u}=1$. This is actually the role of the extra coordinate $w$, whose equation of motion imposes that $\dot{u}$ is indeed constant on-shell. 

Let us have a closer look at the lifted metric and derive the equations of motion of the lifted Lagrangian.
By construction, the lifted metric is very similar to the pp-wave ansatz, with $\partial_w$ and $\partial_u$ both null Killing directions. 
The classical trajectories are geodesics of the lifted metric written in terms of the affine parameter  $\lambda$. The lifted lapse $\N$ imposes that we focus on null geodesics and ensures the invariance of the lifted action under $\lambda$-reparametrization.

We  compute the canonical momenta,
\be
p_u = \frac{\delta \L}{\delta \dot{u}}
=
\frac{\left( \dot{w} - 2V(\chi) \dot{u}\right)}{\N} 
\;, \qquad
p_w = \frac{\delta \L}{\delta \dot{w}}
 = \frac{\dot{u}}{\N}
 \;, \qquad
 p_a =  \frac{\delta \L}{\delta \dot{\chi}^a} = \frac{1}{\N} g_{ab} \dot{\chi}^b
 \,,
\ee
and get the Hamiltonian governing the geodesic motion on the lift,
\begin{align}
\H
=
\L
=
\N \left[ 
p_u p_w +   \frac{1}{2} g^{ab} (\chi)p_{a} p_{b}  + {V(\chi)} p^2_w 
\right]
\,.
\end{align}
Stationarity with respect to  variation of the lifted lapse $\N$ imposes the condition that the vanishing Hamiltonian, $\H=0$.
This restricts the dynamics to null geodesics on the lifted field space.
The remaining  equations of motion read
\begin{align}
\label{condu}
&\rd_{\lambda}{p}_u = \{ p_u, \H\} = 0 \;, \qquad   \rd_{\lambda}{u} = \{ u, \H\} = \N p_w\,,
\\
\label{condw}
&\rd_{\lambda}{p}_w = \{ p_u, \H\} = 0  \;, \qquad \rd_{\lambda}{w} = \{ w, \H\} = \N (p_u+2p_w V)\,,
\end{align}
\be
\rd_{\lambda}{\chi}^{a} = \{\chi^{a}, \H \} =   \N g^{ab}p_b
\;, \qquad
\rd_{\lambda}{p}_a = \{ p_a, \H\} = -   \N \left[ \f12p_{b}p_{c} \partial_a g^{bc} + {p^2_w} \partial_a V \right]
\,.
\ee
%
Let us integrate these equations of motion for the null coordinates $(u,w)$ and their momenta to show the lifted system is equivalent to the initial system \eqref{gen:original}.
First, the coordinate $w$ is a cyclic variable, i.e. its momentum $p_w$ is a constant of motion. For the sake of simplicity, we can set $p_{w}=1$. Then the lifted Hamiltonian reads:
\be
\H
\,\underset{p_{w}=1}=\,
\N\,(p_{u}+h)
\,,\qquad
h=
\frac{1}{2} g^{ab} (\chi)p_{a} p_{b}  + {V(\chi)} \,,
\ee
where $h$ is the Hamiltonian of the original action.
Thus, imposing that the lifted Hamiltonian vanishes $\H=0$ amounts to identifying the $u$-momentum to the Hamiltonian of the initial system (up to a sign switch):
\be
p_{u}
\,\underset{\H=0}=\,-h
\ee
This means that the variable $u$ can be identified to the original time coordinate $\tau$. In particular, we can deparametrize the equations of motion of the lifted system in terms of $u$ and recover the equations of motion of the original system:
\be
\f{\rd \chi^{a}}{\rd u}
=
\f{\rd_{\lambda} \chi^{a}}{\rd_{\lambda} u}
\,\underset{p_{w}=1}=\,
g^{ab}p_{b}
\,,\qquad
\f{\rd p_{a}}{\rd u}
=
\f{\rd_{\lambda} p_{a}}{\rd_{\lambda} u}
\,\underset{p_{w}=1}=\,
-  \left[ \f12p_{b}p_{c} \partial_a g^{bc} + \partial_a V \right]\,,
\ee 
where we recognize the Hamilton equations of motion of the original system \eqref{gen:original} for the geodesics of the metric $g$.
This shows that the geodesic equation for the massless test particle on the lift indeed reproduces the  equations of the initial mechanical system. Finally, the equation of the $w$-coordinate can be recast as
\be
\rd_{\lambda} w 
= \N ( - h +2 V) \; \underset{p_{w}=1} = - \N \left(\frac{1}{2} g^{ab} p_a p_b - V \right) = - \N \cL 
\ee
When $V(\chi) \neq0$, this equation has to be integrated case by case. However, for a free system, one has $V=0$ and thus $\cL = - h$. Since $h$ is conserved, the evolution of $w$ is $w(\lambda) =  \lambda h$.

\medskip

Now that we are back to a system of null geodesics, we know that conformal Killing vectors will give symmetries of the system. More precisely, writing $\G_{AB}$ for the lifted metric,
\be
\G_{uu}=-2V\,,\qquad
\G_{uw}=1\,,\qquad
\G_{ab}=g_{ab}\,
\ee
we consider a conformal Killing vector,
\be
X=X^{A}\pp_{A}=X^{u}\pp_{u}+ X^{w}\pp_{w}+\xi^{a}\pp_{a}\,,
\ee
thus satisfying the Lie derivative equation $\cL_{X}\G=\vphi\,\G$ or explicitly, if we distinguish the null coordinates $(u,w)$ from the original coordinates $\chi^{a}$:
\be
\label{gen:liftCKV}
\begin{array}{lcl}
\vphi &=& \pp_{u}X^{u}+\pp_{w}X^{w}-2V\pp_{w}X^{u}\,,
\\
\vphi V &=& 2V\pp_{u}X^{u}-\pp_{u}X^{w}+\xi^{a}\pp_{a}V\,,
\\
0&=&\pp_{a}X^{w}-2V\pp_{a}X^{u}+g_{ab}\pp_{u}\xi^{b}
\\
0&=&\pp_{a}X^{u}+g_{ab}\pp_{w}\xi^{b}\,,
\\
\vphi\,g_{ab}&=&\cL_{\xi}g_{ab}\,.
\end{array}
\ee
Let us underline that, although the last equation, giving the projection of the conformal Lie derivative onto the original space, strictly reads as before $\cL_{\xi}g=\vphi g$, now the vector components $\xi^{a}$ not only depend on the original coordinates $\chi^{a}$ but can also depend on $u$ and $w$.
We know that the scalar product $X^{A}p_{A}$ is a constant of motion along all null geodesics. The next step consists in translating this conserved quantity at the level of the null geodesic on the lift to a conserved quantity for the initial mechanical system. This can be done using the equations of motion for the coordinates $(u,w)$ and their momenta $(p_u, p_w)$. These equations of motion will be called the projection rules hereafter.

Let us first focus on free systems for which the potential vanishes, i.e. $V= 0$. In that case, the lift metric is symmetric w.r.t to the change $u\leftrightarrow w$ since they only appear in the cross term $\rd u \rd w$. One can therefore first look for all CKV such that $\pp_{w}X=0$ and once the solutions have been obtained, construct the additional solutions by performing the change $u\leftrightarrow w$. However, these vector fields will not leave invariant the key structure of the lift, which is the covariantly constant null vector $\partial_w$. It is therefore usual to split between those CKV which will leave invariant $\partial_w$, and the one will not Lie commute with this vector\footnote{They are also known as chrono-projective vector fields. See \cite{Zhang:2019gdm} for more details.}. Focusing on the former CKV $X^A \partial_A$, i.e, such that $\partial_w X$, and the associated conserved quantity $X^A p_A$, one can translate this lifted Poisson bracket $\{X^{A}p_{A},\H\}_{ED}=-\vphi\H\heq0$ back to the original system, using the mapping derived above from the equations of motion in $(u,w)$,
\be
\label{projj}
p_{w}=1\,,\qquad
p_{u}=-h\,,\qquad
u=\tau\,, \qquad 
w = \tau h 
\ee
we introduce the corresponding observable:
\be
\cO_{X}\equiv X^{w}-hX^{u}+\xi^{a}p_{a}\bigg{|}_{u=\tau}
\,.
\ee
A straightforward calculation allows to check that this is indeed a constant of motion for the original system:
\be
\left|
\begin{array}{l}
\pp_{w}X=0\\
\cL_{X}\G=\vphi\,\G
\end{array}
\right.
\quad\Longrightarrow\quad
\rd_{\tau}\cO_{X}=\pp_{\tau}\cO_{X}+\{\cO_{X},h\}=0
\,.
\ee
A key remark is that this method produces, as desired, evolving constants of motion, i.e. conserved charges which explicitly depend on time, which was not possible before. This follows directly the $u$-dependence of the Killing vector field $X$.
%
For free systems, the lifted metric is the straightforward embedding of the original $d$-dimensional metric into a $(d+2)$-dimensional metric by adding a mere $\rd u \rd w$ contribution for the new null coordinates. Despite being straightforward, the charges derived from the Eisenhart-Duval lift CKVs are non-trivial and explicitly time-dependent and thus come on top of  the ones descending from the CKVs of the original (super-)metric.

Now, one can also apply this procedure to system with a non-vanishing potential $V:=V(u, \chi^a)$\footnote{The potential cannot depends on the $w$-coordinate as it would breaks the killing isometry under the vector $\partial_w$.}. In that case, one looses the symmetry under the change $u\leftrightarrow w$ and the doubling of CKV (and the associated charges) one has for free systems. Nevertheless, the construction of conserved charges and evolving constants of motion (i.e. explicitly time-dependent charges) goes exactly along the same line. Concrete examples of such case have been worked out in \cite{Cariglia:2016oft, Cariglia:2018mos, Dhasmana:2021qvw}. See also \cite{Achour:2022syr} for a recent application of this case to black hole perturbation theory.

We would like to conclude this discussion with the remark that it is tempting to interpret free systems, being themselves null geodesic systems, as Eisenhart-Duval lifts. Actually realizing them as Eisenhart-Duval lifts would mean to be able to recast the $d$-dimensional (super-)metric $g_{ab}$ in the lifted ansatz \eqref{gen:EDmetric}, that is:
\be
g_{ab}\rd \chi^{a}\rd\chi^{b}=2\rd U \rd W -2 \Upsilon(U,\zeta) \rd U^{2} +g^{red}_{\alpha\beta}\rd \zeta^{\alpha}\rd \zeta^{\beta}\,,
\ee
similarly to the pp-wave ansatz, in terms of a pair of null coordinates $(U,W)$, a $(d-2)$-dimensional reduced metric $g^{red}$ and an effective potential $\Upsilon$. As we will see later, it turns out that the 2d mini-superspace for  FRW cosmology can be recast exactly in those terms. We leave the general case for future investigation, but nevertheless point out that such a reformulation of relativistic free systems as Eisenhart-Duval lifts begs the intriguing question of the possibility of a general $d\rightarrow (d-2)$ deparametrization of the relativistic models.

\subsection{On conformal Killing vectors}

\label{sec14}


The strategy that we shall adopt in the next sections to find the dynamical symmetries of a given mini-superspace is to write down its super-metric and its Eisenhart-Duval lift and identify all their CKVs. This raises two basic questions:
(i) What is the maximal number of (linearly independent) CKVs (resp. KVs) for a given curved manifold of dimension $d$ ? (ii) Under which condition a given curved manifold of dimension $d$ is maximally symmetric, i.e. have the maximal number of CKVs ?

These questions have been thoroughly studied in the general relativity literature and we quickly overview known results. The interested reader can refer to \cite{Batista:2017wcm} for detailed proofs.  The maximal number of Killing vectors (KVs)  for a $d$ dimensionl geometry is given by
\be
n_{KV} = \frac{d(d+1)}{2} \,,
\ee
while the maximal number of CKVs that a $d$-dimensional curved manifold can admit is given by
\be
n_{CKV} = \frac{(d+1)(d+2)}{2} \;, \qquad \text{for} \qquad d \geqslant 3
\,.
\ee
The $d=2$-dimensional case  is special since such a geometry can admit at most $n_{KV}=3$ Kiling vectors  while it can admit an infinity of linearly independent conformal Killing vectors.

Finally, a $d$-dimensional curved background is maximally symmetric, i.e. has the maximal number of CKVs, if and only if its Weyl and Cotton tensors vanish identically.
Notice that the Cotton tensor is defined only for $d \geqslant 3$ while the Weyl  tensor always vanishes in $d=2$ and $d=3$ dimensions. Therefore, in $d=3$ dimensions, we only have to check for the Cotton tensor to vanish or not in order to determine whether the geometry is maximally symmetric.

Let us stress that these properties further underline the differences between the superspace and the Eisenhart-Duval lift. Indeed, for a system with $d$ degrees of freedom, the former is $d$-dimensional while the latter is a higher  $d+2$-dimensional manifold. It follows that their curvature properties and thus the maximal number of CKVs they can have can be radically different. By construction, this reflects on the underlying symmetries of the mechanical system which stands as their shadow. Perhaps the most striking example is for a system with two degrees of freedom, such as the flat  FRW cosmology filled with a massless scalar field or the Schwarzschild black hole mechanics. In  those cases, the associated field space is a $2$-dimensional background which is thus conformally flat. Therefore, it possesses an infinite set of linearly independent CKVs. On the contrary, the Eisenhart-Duval lift of these systems is a $4$-dimensional background which possesses at most fifteen linearly independent CKVs. Hence, some symmetries of the field space can not have their counterpart in the ED lift and vice-versa. Thus identifying the underlying symmetries of a given mechanical system via the CKVs of its superspace or its Eisenhart-Duval lift should be considered as two distinct approaches which can reveal different sets of  symmetries.

\medskip

In the following sections, we study in details two gravitational mini-superspaces:
(i) the  FRW cosmological mini-superspace for general relativity coupled to a homogeneous isotropic scalar field and
(ii) the Schwarszchild mechanics for gauge-fixed spherically symmetric space-times formulated as Kantowski-Sachs metrics.
%

\section{Flat  FRW cosmology}
\label{sec: FRW}

In this section, we present the dynamical symmetries of the simplest cosmological model consisting in the flat  FRW geometry filled with a massless scalar field. This model was initially studied in \cite{BenAchour:2019ufa, Achour:2021lqq, BenAchour:2020njq}.

\subsection{Action and phase space}

Consider the line element given by
\be
\rd s^2 = - N^2(t) \rd t^2 + a^2(t) \delta_{ij} \rd x^i \rd x^j
\ee
where $N$ is the lapse and $a$ the scale factor.
The reduced action of the Einstein-Scalar system is given by
\begin{align}
S[g, \phi] & = \int_{\M} \rd^4 x \sqrt{|g|} \left[ \frac{\R}{16\pi G} - \frac{1}{2} g^{\mu\nu} \phi_{\mu} \phi_{\nu}  \right]   \\
& = V_0 \int \rd t \left[ \frac{a^3}{2N} \dot{\phi}^2 - \frac{3}{8\pi G} \frac{a\dot{a}^2}{N}  + \frac{3}{8\pi G} \frac{\rd}{\rd t} \left( \frac{a^2 \dot{a}}{N}\right)\right] 
\end{align}
where $V_0$ is the fiducial volume of the cell on which we restrict the spatial integration. Introducing the volume $v = a^3$, the action becomes
\be
\label{acvol}
S[N, v, \phi ; t] = \frac{c\ell_P}{2} \int \rd t  \left[ \ell_P^2 v \frac{\dot{\phi}^2}{N} - \frac{\dot{v}^2}{ N v} +  \frac{\rd}{\rd t} \left( \frac{3\dot{v}}{N}\right)\right]
\ee
where $\ell_P = \sqrt{12\pi G}$ is the Planck length and we have introduced the dimensionless constant 
\be
\label{c}
c= \frac{V_0}{\ell^3_P}
\ee
which encodes the ratio between the IR and UV cut-offs of our symmetry-reduced system. At this stage, we can omit the total derivative term in the action since it will not modify the Friedmann dynamics nor its symmetries.

In order to discuss the dynamical symmetries of this cosmological system, let us slightly simplify the  FRW action (\ref{acvol}) by introducing the field redefinition
\be
\label{z}
z = \sqrt{v}
\ee
such that the  FRW action becomes
\be
\label{flr}
S[N, z, \phi; t] =  \ell_p c\int \rd t \frac{1}{2N} \left(  \ell_P^2z^2 \dot{\phi}^2- 4 \dot{z}^2 \right)
\ee
It corresponds to a free system of the form (\ref{freeac}) with $\chi^a = \{ \phi, z\}$ where the super-metric is given by $g_{\phi\phi} = \ell^2_P z^2$ and $g_{zz} = -4$. This field redefinition removes the denominator in the gravitational kinetic term which will be useful when computing the finite symmetry transformations.
With this new variable, the momenta and the hamiltonian take the following form
\be
 \left|
    \begin{array}{l}
         p   = - 4  \ell_p cN^{-1}\dot{z} \,,  \\
         \pi = c \ell_P^3 N^{-1}z^2 \dot{\phi} \,, 
    \end{array}
\right. 
\qquad H[N] = N h =  \frac{N}{2c\ell_P^3} \left[ \frac{\pi^2}{z^2} - \frac{\ell_P^2 p^2}{4}\right]
\ee
and the symplectic structure reads
\be
\{ z, p\} =\{ \phi, \pi\} =1
\ee
The equations of motion governing the cosmological dynamics read
\begin{align}
\frac{\dot{z}}{N} = - \frac{p}{4c\ell_P} \;, \qquad \frac{\dot{p}}{N} = \frac{\pi^2}{c\ell^3_P z^3} \;, \qquad \frac{\dot{\phi}}{N} = \frac{\pi}{c\ell_P^3 z^2} \;, \qquad \frac{\dot{\pi}}{N} = 0
\end{align}
supplemented with the constraint $h=0$ enforced by the non-dynamical lapse field. Having reviewed the phase space of the system, we can now present its hidden symmetries and discuss the associated algebra of observables.
As we shall see, this simple cosmological system can be algebraically characterized by different sets of observables.

\subsection{Schr\"{o}dinger observables}

In this section, we present a set of non-independent observables which form a two-dimensional centrally extended Schr\"{o}dinger algebra for the gauge-fixed  FRW system. We discuss the symmetries they generate, the role of the Casimirs and algebraically solve the cosmological dynamics from the knowledge of the charges. The derivation of these Schr\"{o}dinger observables using the Eisenhart-Duval lift is discussed in the next section. The explicit derivation of these generators can be found in Appendix~\ref{app:A}.

\subsubsection{Charge algebra}

\label{sec-Sch}
Consider the first three conserved charges defined by
\begin{align}
\label{sl}
 \left|
 \begin{array}{l}
         Q_{+} = \ell_P h \,,  \\
         Q_0  = \frac{1}{2} z p + \tau h \,, \\
         Q_{-} = c z^2 - \frac{1}{2\ell_P} \left(  \tau z p + \tau^2 h\right)
    \end{array}
\right.
\end{align}
They correspond to the charges already discussed at length in \cite{BenAchour:2019ufa, Achour:2021lqq, BenAchour:2020njq} which form an $\sl(2,\mathbb{R})$ algebra. Consider now the five additional charges
\begin{align}
\label{pb}
         P_{\pm}  = e^{\mp \ell_P \phi/2} \left[  \frac{p}{2}  \pm \frac{\pi}{\ell_P z}  \right] \,,  \qquad 
         B_{\pm}  = e^{\mp \ell_P \phi/2} \left[  2 c  z - \frac{\tau}{\ell_P}  \left( \frac{p}{2} \pm \frac{\pi}{ \ell_P z}   \right) \right]\,,  \qquad J = \frac{\pi}{\ell_P}\,.
\end{align}
The charges are written such that they are dimensionless, i.e. $[Q] = [P] = [B] =[J] = 1$.
It is straightforward to check that these charges are indeed conserved. In particular, while the charges $P$ and $J$ are strong Dirac observables, the charges $Q$ and $B$ depend explicitly on time and play the role of evolving constants of motion which generate dynamical symmetries of the system, i.e. they satisfy (\ref{evolv}).
Altogether, they form the following 9-dimensional Lie algebra
\begin{align}
\begin{array}{c}
\begin{array}{l}
\left\{Q_+,Q_-\right\}=Q_0\;,\qquad\qquad
\left\{Q_0,Q_\pm\right\}=\pm Q_\pm\;,
\end{array}\\ \\
\begin{array}{l}
\left\{Q_-,P_\pm\right\}=\frac{1}{2}B_\pm\;,\\
\left\{Q_+,B_\pm\right\}=P_\pm\;,
\end{array}\qquad
\begin{array}{l}
\left\{Q_0,P_\pm\right\}=+\frac{1}{2}P_\pm\;,\\
\left\{Q_0,B_\pm\right\}=-\frac{1}{2}B_\pm\;,
\end{array}\qquad
\begin{array}{l}
\left\{J,P_\pm\right\}=\pm \frac{1}{2}P_\pm\;,\\
\left\{J,B_\pm\right\}=\pm \frac{1}{2}B_\pm\;,
\end{array}
\end{array}
\end{align}
while
\be
 \{ B_{\mp}, P_{\pm}\}  = \nb
\ee
where the central extension is given by $\nb = 2 c $. 

The observables $Q_0$ and $Q_{\pm}$ form the $\sl(2,\mathbb{R})$ sector and generate M\"{o}bius transformations.
The generators $P_{\pm}$ and $B_{\pm}$ give rise to the two $\mathbb{R}^{2}$ subalgebras. They generate respectively the translations and galilean boosts.
The generator $J$ generates a boost in both the $P_{\pm}$ plane and the $B_{\pm}$ plane. This might be a surprising feature since $J$ is simply the dimensionless scalar field momentum, but it reflects the peculiar exponential factor $e^{\mp \ell_P \phi/2} $ appearing in the $P_{\pm}$'s and the $B_{\pm}$'s. We underline that this is not a boost at all in space or space-time but a boost in field space. In the standard Schr\"{o}dinger algebra, $J$ would be a compact rotation generator associated to an $\so(2)$ sector, but in our case, the 2d field space on which it acts has the signature $\text{diag}(-,+)$, turning it into a boost generator. It follows that the above algebra can be recognized as a Schr\"{o}dinger algebra 
\be
\sh(2) =( \sl(2,\mathbb{R})\times \so(1,1) )  \ltimes( \mathbb{R}^2 \times \mathbb{R}^2)
\ee
with a rotational sector replaced by a boost.
In the following, we shall keep using the name Schr\"{o}dinger algebra despite this difference.

There are three independent Casimirs associated to this 2d Schr\"{o}dinger algebra \cite{Alshammari:2017jky}. The first one is the central extension, i.e. $\C_1 = n = 2c$, while the two other Casimirs are given by
\begin{align}
\label{secCaz}
\C_2  & =  P_{+}B_{-}-P_{-}B_{+}-2 n J
=0 \\
\C_{3}
&=
Q_{0}^{2}-Q_{+}Q_{-}- J^{2}
- B_{+}B_{-}Q_{+}- P_{+}P_{-}Q_{-} \nn \\
& \;\;\;  - (B_{-}P_{+}+B_{+}P_{-})Q_{0}
+(B_{-}P_{+}-B_{+}P_{-})J
=
0
\,.\nn
\end{align}
Anticipating on the quantization of this cosmological model, we expect that the last two Casimirs no longer vanish in the quantum theory. Their non-vanishing values encodes the presence of the additional degrees of freedom at the quantum level which correspond to the moments of the wave function. Their knowledge therefore provides an algebraic characterization of the deviation from classical to quantum cosmology.

 \subsubsection{Symmetry transformations}

\label{sym}

Consider the action (\ref{flr}) for the flat  FRW cosmology, $S = \f{c \ell_P}2 \int N^{-1}(  \ell_P^2z^2 \dot{\phi}^2- 4 \dot{z}^2 )\rd t$.
Here, we give  the finite symmetry transformations generated by the conserved charges identified above, and check explicitly the invariance of action.

The translations and Galilean boosts generated respectively by the conserved charges $P_{\pm}$ and $B_{\pm}$ can be compactly written as
\begin{align}
\tau \rightarrow \tilde{\tau} & = \tau\\
\label{bbb}
z \rightarrow \tilde{z} (\tilde{\tau})& = z(\tau) + \frac{\xi(\tau)}{2}  e^{\epsilon \ell_p \phi/2} \\
\label{bbbb}
\phi \rightarrow \tilde{\phi}(\tilde{\tau}) & =  \phi(\tau) -  \frac{\epsilon}{\ell_P z}  \xi(\tau)e^{\epsilon \ell_P \phi/2}
\end{align}
where $\xi(\tau)$ is an arbitrary function and $\epsilon = \pm 1$. Considering the infinitesimal version of  these transformations, the action transforms as
\begin{align}
\delta_{\xi} S = 2 c\ell_P  \int \rd \tau \left\{  \frac{\rd}{\rd \tau} \left(  \dot{\xi} z e^{\epsilon \ell_P \phi/2} \right) - \ddot{\xi} z e^{\epsilon \ell_P \phi/2}+ \O(\xi^2)\right\}
\end{align}
which shows that the above transformations are indeed Noether symmetries provided $\ddot{\xi} =0$. This implies that
\be
\xi(\tau) = \lambda \tau + \eta
\ee
where $(\lambda, \eta)$ are the two parameters of the symmetry.
Let us derive the associated conserved charges using directly the Noether theorem. The infinitesimal variation of the fields are given by
\begin{align}
\delta_{\xi, \epsilon} z = \frac{\xi}{2}  e^{\epsilon \ell_P \phi/2 } \qquad \delta_{\xi,  \epsilon} \phi = - \frac{\epsilon}{\ell_P z} \xi e^{\epsilon \ell_P \phi/2 }
\end{align}
The Noether charges read $q_{\xi} = \Theta (z, \phi, \delta_{\xi,  \epsilon} z, \delta_{\xi,  \epsilon} \phi) - \delta_{\xi,  \epsilon} L$ where $\Theta$  is the symplectic potential and $L$ is the lagrangian. Explicitly, we have
\begin{align}
\Theta (z, \phi, \delta_{\xi,  \epsilon} z, \delta_{\xi,  \epsilon} \phi) & = c \ell^3_P z^2 \dot{\phi} \; \delta_{\xi,  \epsilon} \phi - 4 c \ell_p \dot{z} \;  \delta_{\xi,  \epsilon} z  = e^{\epsilon \ell_P \phi/2 } \left[ \frac{p}{2} - \epsilon \frac{\pi}{\ell_p z}\right] \\
\delta_{\xi} L & = 2 c \ell_p  \dot{\xi} z e^{\epsilon \ell_P \phi/2 }
\end{align}
Combining them, the Noether charge takes the form
\be
q_{\xi,  \epsilon} = -  \ell_P e^{\epsilon \ell_P \phi/2 } \left[ 2 c  z \dot{\xi} - \frac{\xi}{\ell_P} \left( \frac{p}{2} - \epsilon \frac{\pi}{\ell_p z}\right) \right]
\ee
Using the expression for $\xi$ derived above, i.e. $\xi= \lambda \tau + \eta$, and using that $\epsilon = \pm 1$, one finds that
\begin{align}
q_{\eta, \pm} & =   \eta \;   e^{\pm \ell_P \phi/2 } \left( \frac{p}{2} \mp \frac{\pi}{\ell_p z}\right)  = \eta P_{\mp} \\
q_{\lambda, \pm} & =  - \lambda \ell_P \;  e^{\pm \ell_P \phi/2 } \left[ 2 c  z  - \frac{\tau}{\ell_P} \left( \frac{p}{2} \mp \frac{\pi}{\ell_p z}\right) \right] = - \lambda \ell_P B_{\mp}
\end{align}
One recovers as expected the charges introduced in the previous section which were constructed by the Eisenhart-Duval lift method. Finally, the simplest symmetry of the system consists in translation in the $\phi$-direction in field space, i.e. $\delta \phi = \ell_P \alpha$ and $\delta z =0$ with $\alpha$ a dimensionless constant. One has $\delta_{\alpha} L =0$ and the Noether charge reads $q_{\alpha} = \Theta (z, \phi, \delta_{\alpha} z, \delta_{\alpha} \phi) = \ell_P \alpha \pi = \alpha J$, reproducing the generator of boost in field space. 

The remaining symmetries generated by the charges $Q_{\pm}$ and $Q_0$ have been studied at length in previous works \cite{Pioline:2002qz, BenAchour:2019ufa, BenAchour:2020njq, Achour:2021lqq}. They correspond to the following conformal transformations
\begin{align}
\tau \rightarrow \tilde{\tau} & = f(\tau)\\
z \rightarrow \tilde{z} (\tilde{\tau})& = \dot{f}^{1/2} z(\tau)\\
\phi \rightarrow \tilde{\phi}(\tilde{\tau}) & =  \phi(\tau) 
\end{align}
At the infinitesimal level, i.e. for $f(\tau) \simeq \tau + \chi(\tau)$, the action transforms as
\begin{align}
\delta_{\chi} S = c\ell_P \int \rd \tau \left\{ \dddot{\chi} z^2 + \frac{\rd}{\rd \tau} \left[ \frac{\chi}{2} \left(4\dot{z}^2 - \ell_P^2 z^2 \dot{\phi}^2 \right) - \ddot{\chi} z^2\right] + \O(\chi^2)\right\}
\end{align}
This variation is a Noether symmetry provided $\dddot{\chi} =0$ which imposes
\be
\chi(\tau) = \alpha_{-} + \alpha_{0} \tau + \alpha_{+} \tau^2
\ee
where $(\alpha_{\pm}, \alpha_0)$ are real constants parametrizing this symmetry. We recover the symmetries presented initially in \cite{BenAchour:2019ufa} for the flat  FRW cosmological model and extended to the (A)dS cosmology and the closed and open universes in \cite{Achour:2021lqq}. We do not reproduce the computation of the Noether charges via the Noether theorem since this computation has already been presented in details for the charges $Q_{\pm}$ and $Q_0$ in several previous works. See for example Section~(2.1.3) in \cite{BenAchour:2019ufa} or Section~(4) in \cite{Achour:2021lqq}.

An interesting question which goes beyond the scope of this work is to fully characterize the role of the gauge-fixing in the realization of this Schr\"{o}dinger symmetry. This point was discussed in detailed in \cite{Achour:2021lqq} for the restricted SL$(2,\mathbb{R})$ symmetry. Let us point that the present symmetries should not be regarded as the standard diffeomorphisms as the dynamical gravitational field, the scale factor $z$ in the present case, does not transform as a scalar quantity but as a primary field with a fixed conformal weight. In that sense, these transformations are new non-standard symmetries of the (homogeneous) gravitational field.

\subsubsection{Integration of the dynamics}

Now, let us show that the knowledge of this charge algebra allows one to algebraically solve the  FRW dynamics. One can distinguish between the information encoded on one hand in the strong Dirac observables, such as $J$ and  $P_\pm$, and on the other hand in the evolving constants of motion, i.e. the charges which depend explicitly on the time coordinate.

The first charges allows one to solve for the deparametrized dynamics. For instance,  the translational charges
%
$P_{\pm}=e^{\mp \ell_P \phi/2} \left[  \frac{p}{2}  \pm \frac{\pi}{\ell_P z}  \right] $
allow to directly solve the deparametrized trajectories $v(\phi)$ which read
\be
v(\phi )\heq V^2_{\pm} e^{\pm \ell_P \phi } \;, \qquad \text{with} \qquad V_{\pm} = \frac{Q_0\pm J}{P_{\pm}}
\ee
These conserved charges are to be understood as relational observables, in the sense that they encode (time independent) relations between  the dynamical variables. These relations tell how variables varies in terms of the others. If one has a complete set of such observables, then one can deparametrize the system and tell express the evolution of all dynamical variables in terms of any single variable of the system. Identifying those relational observables are thus exactly equivalent to solving for the timeless deparametrized trajectories of the system.

On the contrary, the evolving constants of motion allow one to algebraically determine the physical trajectories w.r.t. the time coordinate $\tau$. Remember that on-shell, the hamiltonian of the system has to vanish which reflects the relativistic nature of the system. We have therefore $Q_{+}\heq 0$. Combining the expressions of $Q_{-}$ and $Q_{0}$ allows one to solve for the physical trajectory $v(\tau)$ which reads
\begin{align}
\label{ FRWvol}
 v(\tau) \coloneqq  z^2(\tau) \heq \frac{2}{N} \left[ Q_0 \frac{\tau}{\ell_P} + Q_{-} \right] \;, 
\end{align}
and we recover the linear growth of the volume in presence of a scalar field driving the cosmic expansion. Notice that the sign of $Q_0$ determines the expanding/contracting behavior of the universe.
This linear evolution of the spatial volume clearly leads to a finite time crash, either in the past if $Q_{0}$ is positive or in the future if $Q_{0}$ is negative. This singularity is a standard feature of flat  FRW cosmology.

In order to solve the profile of the scalar field, notice that 
\be
Q_{+}= -\frac{1}{2c} P_{+} P_{-} \heq 0\,,
\ee
so that  physical trajectories correspond to either $P_{-}\simeq 0$ and $P_{+}\neq 0$, or $P_{+}\simeq 0$ and $P_{-}\neq 0$.  Choosing the first case, we can use the profile of $z(\tau)$ and the charge $P_{+}$ to find
\be
\phi (\tau) \heq   - \frac{1}{\ell_P} \log{\left[ \frac{2}{N} \left( Q_0 \frac{\tau}{\ell_P} + Q_{-}\right)\right]} + \frac{2}{\ell_P} \log{\left[ \frac{Q_0 + J}{P_{+}} \right]}\;, 
\ee
Thus, as expected, the conserved charges provide a complete set of observables which fully encode the  FRW dynamics.  As a result, one can algebraically solve the cosmological dynamics solely based on the knowledge of the Schr\"{o}dinger cosmological observables. Now let us  present how to derive the different charges introduced in this section from a geometrical point of view.

\subsection{Symmetry generators from the Eisenhart-Duval lift}

\label{Tech}

In this section, we show that the Schr\"{o}dinger observables  can be derived from conformal Killing vector fields of the Eisenhart-Duval (ED) lift of the  FRW system. The details are presented in Appendix~\ref{app:A}.

By construction, the ED lift of the flat  FRW cosmology has the following line element
\be
\label{FLIFT}
\rd s^2 = g_{AB} \rd X^{A} \rd X^{B} = 2\rd u \rd w +  \ell_p c\left(  \ell_P^2 z^2 \rd \phi^2 - 4 \rd z^2\right)
\ee
Let us denote $N^A\partial_A = \partial_w$ the covariant constant null vector. It is straightforward to show that the Weyl and Cotton tensor associated to this four dimensional geometry vanish. Therefore, it is conformally flat and thus maximally symmetric. It follows that it possesses the maximal number of CKVs, i.e. fifteen, which form the $\so(4,2)$ Lie algebra. Let us stress here that we refer to the conformal flatness of the lift metric (\ref{FLIFT}), and not to the conformal flatness of the FRW spacetime metric. As already pointed out in the first section, this set of CKVs can be organized with respect to their action on the covariantly constant null Killing field $N^A\partial_A = \partial_w$. 

Consider firstly the set of CKVs which commute with $\partial_w$. There are eight such CKVs. The first three are given by
\begin{align}
Q^{A}_{+} \partial_A & =- c  \partial_u\\
Q^{A}_{0} \partial_{A}&  = \frac{1}{2} z \partial_z + u \partial_u \\
\label{speconf}
Q^{A}_{-} \partial_{A} & = c\ell_P z^2 \partial_w  + \frac{1}{2} \left(  u z \partial_z  +  u^2 \partial_u \right)
\end{align}
while the five remaining vectors fields are given by
\begin{align}
P^A_{\pm} \partial_A & = e^{\mp \ell_P \phi/2} \left[ \frac{\ell_P}{2} \partial_z \pm \frac{1}{z} \partial_{\phi}\right] \;, \\
B^A_{\pm} \partial_A & = e^{\mp \ell_P \phi/2} \left[ 2 c \ell^2_P z \partial_w + u \left( \frac{\ell_P}{2} \partial_z \pm \frac{1}{z} \partial_{\phi}\right) \right]\;, \\
J^A \partial_A & = \partial_{\phi}
\end{align}
Together with $N^A\partial_A$, these eight vectors fields form the 2-dimensional Schr\"{o}dinger algebra. Moreover, applying the projections rules reviewed in section~\ref{sec:EDlift} reproduces the Schr\"{o}dinger observables presented in section~\ref{sec-Sch}. These charges are at most quadratic in the momenta. Let us present one example of such projection. Consider the CKV $Q^{A}_{-} \partial_{A}$. The associated conserved charge for the null geodesic motion on the lift is given by
\begin{align}
Q_{-} = Q^A_{-} \; p_A =  c\ell_P z^2 p_w  + \frac{1}{2} \left(  u z p_z  +  u^2 p_u \right)
\end{align}
Imposing the projection rules
\be
p_w = c \;, \qquad p_u = - \frac{h}{c} \;, \qquad u = - c \tau
\ee
one obtains
\be
 Q_{-} =  c\ell_P \left[ c z^2  -  \frac{1}{2\ell_P}  \left(  \tau z p_z  +  \tau^2 h \right) \right]
\ee
which reproduces the third charge in (\ref{sl}) up to an irrelevant global factor $c\ell_P$. Proceeding the same way, one can derive the other Schr\"{o}dinger observables in a straightforward manner.

The remaining CKVs which complete the $\so(4,2)$ algebra do not commute with the constant null vector. Since the metric of the lift is invariant under the switch $u \leftrightarrow w$, we have four dual charges which read
\begin{align}
\tilde{Q}^{A}_{0} \partial_{A}&  = \frac{1}{2} z \partial_z + w \partial_w  \\
\label{speconf}
\tilde{Q}^{A}_{-} \partial_{A} & = c\ell_P z^2 \partial_u  + \frac{1}{2} \left(  w z \partial_z  +  w^2 \partial_w \right) \\
\tilde{B}^A_{\pm} \partial_A & = e^{\mp \ell_P \phi/2} \left[ 2 c \ell^2_P z \partial_u + w \left( \frac{\ell_P}{2} \partial_z \pm \frac{1}{z} \partial_{\phi}\right) \right]\;, 
\end{align}
It is interesting to notice that upon imposing the projection rules (\ref{projj}), the charges $\tilde{Q}_{-}$ and $\tilde{B}_{\pm}$ turn out to be quartic and cubic in the momenta, while $\tilde{Q}_0 = Q_0$ and does not provide a new information on the system. The reason for that is the dependency on the $w$-coordinate of these charges. Indeed, the projection rule for the coordinate $w$ is $w = - \tau h$ where $h$ is the hamiltonian, which is thus quadratic in the momenta. The same can be said when a CKV has a non vanishing component $\xi^u$. This will give rise to a term proportional to $p_u$ for the charge, which upon using the projection rule, i.e. $p_u = -h$, gives a contribution quadratic in the momenta to. Terms cubic in the momenta show up for instance from the second term in $\tilde{Q}_{-}$, i.e. $w z \partial_z$. The associated conserved charge inherits a term of the form $w z p_z $ which upon using the projection rule for $w$, gives $\tau z p_z h $. Such term is thus cubic in the momenta. Finally, the two last CKVs are given by
 \be
Y^A_{\pm} \partial_A = e^{\pm \ell_P \phi/2} \left[ z \left( u \partial_u + w \partial_w \right) + \frac{1}{2c \ell^2_P } \left(2  \ell_p cz^2 + uw \right)\left( \frac{\ell_P}{2}\partial_z \pm\frac{1}{z}\partial_{\phi}\right)\right]
\ee
which are also associated to charges cubic in the momenta.
Those charges are clearly conserved since they are polynomial combinations of the already derived constants of motion with extra Hamiltonian factors in $h$.

It is straightforward to check that these fifteen vectors fields indeed form a $\so(4,2)$ algebra, see appendix~\ref{so42} for the details. We will not look into the finite symmetry transformations that they generate. Indeed, this $\so(4,2)$ algebra of observables plays a similar role as the $\so(4,2)$ conformal symmetries of the quantum 2-body problems, as for the spectral analysis of the Hydrogen atom \cite{Hatom}. Thus the finite transformations do not seem especially relevant at the classical level, although it could be interesting to see how they act on classical trajectories, but they will definitely be relevant to the quantization and spectrum of the quantum theory. Now, let us turn to black hole mechanics.

%

\section{Schwarzschild black hole mechanics}
\label{sec:BH}

In this section, we consider the second example of interest: the Schwarzschild black hole mechanics. This mini-superspace possesses again two degrees of freedom. We show that just as the flat  FRW cosmology filled with a massless scalar field, it enjoys again a set of Schr\"{o}dinger observables which fully encodes the Schwarzschild geometry, both in the exterior and interior regions of the black hole. Previous investigations on the dynamical symmetries of the Schwarzschild mechanics were presented in \cite{Geiller:2020xze, Achour:2021dtj}.

\subsection{Action and phase space}

Consider the spherically symmetric geometry with line element
\be
\label{metric}
\rd s^2 = \epsilon \left(  - N^2(x) \rd x^2 + A^2(x) \rd y^2\right) + \ell^2_s B^2(x) \rd \Omega^2
\ee
where $N(x)$ is the lapse, $\ell_s$ is a fiducial length scale such that the $B$-field is dimensionless and $\epsilon = \pm 1$ is a parameter which will allow us to treat both the interior and the exterior black hole regions at once.
To map the line element onto the Schwarzszchild metric, the coordinate $x$ is the radial coordinate, while the $y$ coordinate is the time coordinate $t$. The sector $\eps=-1$ describes the usual exterior of the Schwarzszchild metric, for which $x$ is space-like and $y$ is time-like and for which we use a time-like foliation of the bulk. The sector $\eps=+1$ describes the interior of the Schwarzszchild black hole, for which $x$ becomes time-like and $y$ space-like and where we use a spacelike foliation of the geometry. See \cite{Achour:2021dtj} for detail.

Now, let us write the action governing its dynamics. The Einstein-Hilbert action reduces to
\begin{align}
 S_{\epsilon}[N, A, B ;x] & = \int \rd x \frac{\R}{2 \ell_P^2}  \\
& = \frac{\ell_0 \ell^2_s}{\epsilon \ell_P^2} \int \rd x \left\{ \frac{\epsilon N A}{\ell^2_s} - \frac{A (B')^2 + 2 B B'A'}{N} + \frac{\rd}{\rd x} \left( \frac{B^2 A' + 2 A B B'}{N}\right)\right\}
\end{align}
where we have introduced the length $\ell_0$ which sets the size of the system in  { the} $y$-direction. At this stage, we introduce a new time coordinate $N \rd x = \N \rd \eta /A $ such that the metric reads
\be
\label{meet}
\rd s^2 = \epsilon \left( - \frac{\N^2(\eta) \rd \eta^2}{A^2(\eta)} + A^2(\eta) \rd y^2\right) + \ell^2_s B^2(\eta) \rd \Omega^2
\ee
This field redefinition allows one to simplify the potential term in the action, such that once we gauge-fixed the $\eta$-reparametrization invariance, i.e. when working with the coordinate $\rd \tau = \N \rd \eta$, the potential term reduces to a simple constant in the mechanical action. Proceeding that way, the symmetry-reduced action becomes
\begin{align}
\label{acSch}
S_\epsilon[N, A, B; \tau]  = \epsilon c\ell_P \int  \rd \tau \left[ \frac{\epsilon}{\ell^2_s} - (A^2 \dot{B}^2 + 2 AB \dot{B}\dot{ A})\right]
\end{align}
In this mechanical action, we have omitted the total derivative term and a dot refers to a derivative w.r.t. the new coordinate $\tau$. Notice also that we have introduced the dimensionless parameter
\be
\label{ccc}
c = \frac{\ell_0 \ell^2_s}{ \ell_P^3}
\,.
\ee
Let us point out that we have two independent IR regulators $\ell_{0}\ne\ell_{s}$, defining integration cut-off respectively in the $y$ and $x$ directions. $\ell_{s}$ is the fiducial radius of the 2-sphere, while $\ell_{0}$ is the interval for the  $y$-coordinate.

We can now solve the equations of motion w.r.t. the coordinate $\rd \tau = \N \rd \eta$. A straightforward computation gives the following profiles for the $A$-field and $B$-field:
\begin{align}
\label{sol}
A^2(\tau) =  -  \frac{\epsilon}{\C^2 \ell^2_s} \frac{\tau- \tau_1}{\tau- \tau_0}\;, \qquad B(\tau) = \C (\tau - \tau_0)
\end{align}
where $\C$ and $(\tau_0, \tau_1)$ are constants of integration. Injecting these solutions in the metric (\ref{meet}), one finds
\begin{align}
\rd s^2 = - \left[\frac{\tau - \tau_1}{\tau - \tau_0} \right] \left[ \rd \left( \frac{y}{\C \ell_s}\right) \right] ^2  + \left[\frac{\tau - \tau_1}{\tau - \tau_0} \right]^{-1} \left[ \rd \left( \C \ell_s \tau\right) \right]^2 + \ell^2_s \C^2 (\tau - \tau_0)^2 \rd \Omega^2 
\end{align}
Performing a translation $\tau \rightarrow \tau + \tau_0$ on the solution, one can rewrite it as
\begin{align}
\rd s^2 = - \left[ 1- \frac{\tau_1 - \tau_0}{\tau } \right] \left[ \rd \left( \frac{y}{\C \ell_s}\right) \right] ^2  + \left[1- \frac{\tau_1 - \tau_0}{\tau } \right]^{-1} \left[ \rd \left( \C \ell_s \tau\right) \right]^2 + (\ell_s \C \tau)^2 \rd \Omega^2 
\end{align}
Finally, rescaling appropriately the coordinates $(y,\tau)$, i.e. $y \rightarrow \C \ell_s$ and $\tau \rightarrow \tau / (\C \ell_s)$, one finds 
\begin{align}
\rd s^2 = - \left[ 1- \frac{\tau_m}{\tau } \right] \rd y^2  + \left[1- \frac{\tau_m}{\tau } \right]^{-1} \rd \tau^2 + \tau^2 \rd \Omega^2 
\end{align}
where we have introduced the quantity $\tau_m$
\be
\tau_m = \C \ell_s (\tau_1 - \tau_0)
\ee
which one identifies to the Schwarzschild mass. When $\tau > \tau_m$, the $y$-coordinate and $\tau$-coordinate play the roles respectively of the time and radius coordinates, while for $\tau < \tau_m$, their role is switched,  The solution corresponds to the Schwarzschild black hole geometry, showing that the simple mechanical action (\ref{acSch}) reproduces indeed the Schwarzschild mechanics.

Now, in order to identify the different relevant families of observables of the black hole mechanics, it will be useful to rewrite the action (\ref{acSch}) in a more appropriate form.
Consider the field redefinition
\be
\label{relAB}
V_1 \coloneqq B^2 \;, \qquad V_2 \coloneqq \frac{A^2 B^2}{2}
\ee
In term of these new fields, the (gauge-fixed) Schwarzschild mechanics (\ref{acSch}) can be recast as
\be
\label{gaugefixSch}
S_{\epsilon} [V_1, V_2 ; \tau]= \epsilon  \ell_p c\int \rd \tau \left[ \frac{\epsilon }{\ell^2_s} +  \frac{V_2 \dot{V}^2_1- 2 V_1 \dot{V}_1 \dot{V}_2}{2V^2_1}  \right]
\ee
The momenta are given by
\be
P_1 = \frac{\epsilon c\ell_P}{V_1^2} \left( V_2 \dot{V}_1 - V_1 \dot{V}_2\right) \;, \qquad P_2 = - \epsilon c\ell_P \frac{\dot{V}_1}{V_1}
\ee
and we have
\be
\dot{V}_1 = - \frac{1}{\epsilon c\ell_P} V_1 P_2 \;, \qquad  \dot{V}_2 = - \frac{1}{\epsilon c\ell_P} \left( V_1 P_1 + V_2 P_2\right)
\ee
The hamiltonian reads 
\be
h = - \frac{1}{\epsilon c\ell_P} \left[ V_1 P_1 P_2 + \frac{1}{2} V_2 P^2_2\right] - \frac{c \ell_P}{ \ell^2_s}
\ee
Because the potential term of the hamiltonian is a mere constant, it is convenient to work with the shifted hamiltonian $\tilde{h} = h + c\ell_P/ \ell^2_{s}$ such that on-shell, one has $\tilde{h} \heq c\ell_P/ \ell^2_{s}$. 
The  equations of motion are given by
\begin{align}
\dot{V}_1 = \{V_1, h \} & = - \frac{1}{\epsilon c\ell_P} V_1 P_2 \;, \qquad \qquad \qquad   \dot{P}_1 = \{ V_1, h\} = \frac{1}{\epsilon c\ell_P} P_1 P_2 \\
\dot{V}_2 = \{V_2, h \} & = - \frac{1}{\epsilon c\ell_P} \left( V_1 P_1+ V_2 P_2\right) \;, \qquad \dot{P}_2 = \{ P_2, h\} = \frac{P^2_2}{2\epsilon c\ell_P} 
\end{align}
Before introducing the Schr\"{o}dinger observables of the system, it is useful to  { first} introduce the following phase space function
\be
\label{gen}
C \coloneqq c\ell_P \{ V_2, h\} = - \epsilon \left( V_1 P_1 + V_2 P_2\right)
\ee
which is a weak Dirac observable of the system satisfying
\be
\{ C, h\} = -\epsilon h
\ee
Having presented the mechanical system which reproduces the Schwarzschild mechanics, we now show that it is equipped with a set of Schr\"{o}dinger observables but also with an infinite tower of $\cw$ observables in a way similar to the  FRW cosmological model.

\subsection{Schr\"{o}dinger observables}

In this section, we present the Schr\"{o}dinger observables of the Schwarzschild mechanics and the associated Casimirs. Then, we show that the Schwarzschild solution can be recovered solely from the knowledge of the conserved charges. 

\subsubsection{Charge algebra}

\label{schsch}

Consider the first three observables  
\begin{align}
Q_{+} & = \tilde{h}   \\
Q_{0} \; & =  \tau \tilde{h} + \epsilon C  \\
Q_{-}  & =  \tau^2 \tilde{h}  + 2\epsilon \tau C - 2 \epsilon  \ell_p c V_2 
\end{align}
where $\tilde{h}$ is the shifted hamiltonian.
They correspond to the conserved charges already discussed in \cite{Geiller:2020xze, Achour:2021dtj}. Next, we obtain the four additional conserved charges given by
\begin{align}
\label{bb}
P_+&=\sqrt{V_1}P_1+\frac{V_2}{2\sqrt{V_1}}P_2 \;, \qquad \ell_P B_+= \frac{\epsilon  \ell_p cV_2}{\sqrt{V_1}}+ \tau P_{+}\\
\label{pp}
P_-&=\sqrt{V_1}P_2\;, \qquad \qquad \qquad \;\;\; \ell_P B_-=2\epsilon c \ell_P\sqrt{V_1}+\tau P_{-}\;, 
\end{align}
to which we can add
\be
J = 2 V_1 P_1.
\ee
Notice that charges $Q$ and $B$ have the dimension of a length, i.e. $[Q]=[B] = L$ while the charges $J$ and $P$ are dimensionless, i.e. $[J]=[P] = 1$.
A straightforward computation shows that these quantities are indeed conserved charges. In particular, the charges $Q_{0}$,  $Q_{-}$ and $B_{\pm}$ are evolving constants of motion and depend explicitly on the $\tau$-coordinate. Therefore, they do not commute with the hamiltonian and change the energy. On the other hand, the other charges are strong Dirac observables of the system. The first three charges form an $\sl(2,\mathbb{R})$ algebra
\be
 \left\{ Q_{+}, Q_{-}\right\} = 2 Q_0\;, \qquad 
\left\{ Q_0, Q_{+} \right\} = - Q_{+}\, ,\qquad 
\left\{ Q_0, Q_{-}\right\} = + Q_{-} 
\ee
while the other brackets read
\be
\begin{array}{l}
\end{array}\qquad 
\begin{array}{l}
\left\{  Q^{\epsilon}_{-}, P_\pm  \right\}=-B^{\epsilon}_\pm\, ,\\
\left\{ Q_{+}, B^{\epsilon}_\pm   \right\}=P_\pm \,,
\end{array} \qquad 
\begin{array}{l}
\left\{ Q^{\epsilon}_{0}, P_\pm   \right\}=-\frac{1}{2}P_\pm\,,\\
\left\{  Q^{\epsilon}_{0}, B_\pm  \right\}=\frac{1}{2}B_\pm\, ,\\
\end{array}\qquad
\begin{array}{l} 
\left\{ J, P_{\pm}\right\} = \pm \; \frac{1}{2} P_{\pm}\,,\\
\left\{ J, B_{\pm}\right\} = \pm \;  \frac{1}{2} B_{\pm}\,.
\end{array}
\ee
Finally, the boosts and translations form the 2d Heisenberg algebra:
\be
\{ B_\mp, P_\pm \} =  n \;, 
\ee
where $\nb = \epsilon c$ is the central extension. We recognize again the two dimensional centrally extended Schr\"{o}dinger algebra $\sh(2)$ which shows that the Schwarzschild mechanics can indeed be equipped with a set of Schr\"{o}dinger observables.  Notice that again, the generator $J$ should not be interpreted as a compact rotation, but as a boost in field space.

Now, let us discuss the interpretation of the Casimirs. For such $\sh(2)$ algebra, there are three independent Casimirs \cite{Alshammari:2017jky}. The first one corresponds to the central extension $\nb$ such that
\be
\C_1 = \nb= \epsilon c =  \frac{\epsilon \ell_0 \ell^2_s}{\ell^3_P}
\ee
The two other Casimirs are again vanishing, i.e. one has
\begin{align}
\label{secCaz}
\C_2  & =  P_{+}B_{-}-P_{-}B_{+}-  n J
=0 \\
\C_{3}
&=
n \left( Q_{0}^{2}-Q_{+}Q_{-}- \frac{1}{4}J^{2} \right) 
- B_{+}B_{-}Q_{+}-  P_{+}P_{-}Q_{-} \nn \\
&\;\;\;  -  (B_{-}P_{+}+B_{+}P_{-})Q_{0}
+\frac{1}{2}(B_{-}P_{+}-B_{+}P_{-})J
=
0
\,.\nn
\end{align}
Just as in the first case, we expect the last two Casimirs be non-vanishing at the quantum level, encoding thus deviation from the classical Schwarzschild geometry.

The Schr\"{o}dinger observables presented in this section can be derived from the conformal isometries of the Eisnehart-Duval lift of the Schwarzschild action, as explained in the first section. Since a first example has been treated in detail in the previous section, we do not reproduce the computation of the CKVs here but we refer the reader to appendix~\ref{app:C} where the expressions of the CKVs are given explicitly. An interesting outcome of this investigation is that the ED lift of the black hole is again a four dimensional conformally flat manifold, such that its CKVs form again an $\so(4,2)$ Lie algebra. The Schr\"{o}dinger charges presented above correspond to the CKVs which commute with the covariantly constant null vector of the ED lift.

At this stage, let us point that an additional symmetry has been found in earlier works \cite{Geiller:2020xze} which, together with the SL$(2,\mathbb{R})$ charges, form an ISO$(2,1)$ symmetry group. This symmetry group turns out to also be a symmetry of the Schwarzschild-(A)dS mechanics \cite{Achour:2021dtj}. Interestingly, the translational charges which build up the Poincar\'e group are quadratic in the momenta $P_2$. For that reason, they cannot be derived by focusing only on the CKVs of the ED lift. Instead, one has to investigate the Killing tensors of the lift to identify its geometrical origin, as discussed in \cite{Cariglia:2014dwa}. A geometrical derivation of these charges and the extension to the Schwarzschild-(A)dS case will be presented elsewhere.

\subsubsection{Symmetry transformations}

We now show that the Schr\"{o}dinger charges presented above generate well defined Noether symmetries of the gauge-fixed Schwarzschild mechanical action (\ref{acSch}). The symmetries induced by the $\sl(2,\mathbb{R})$ charges $Q_{\pm}$ and $Q_0$ have been derived previously in \cite{Geiller:2020xze, Achour:2021dtj} and we shall not reproduce this computation here. The symmetry transformations are given explicitly by
\begin{align}
\tau \rightarrow \tilde{\tau} & = f(\tau) \\
V_1 \rightarrow \tilde{V}_1(\tilde{\tau}) & =  \dot{f}(\tau) V_1(\tau) \\
V_2 \rightarrow \tilde{V}_2(\tilde{\tau}) & =  \dot{f}(\tau) V_2(\tau) 
\end{align}
where the function $f(\tau)$ is given by
\be
f(\tau) = \frac{a\tau + b}{c\tau+d} \;, \qquad \text{with} \qquad ad-bc \neq 0
\ee
The interested reader is referred to \cite{Geiller:2020xze, Achour:2021dtj} for an explicit expression of the transformed action.

Now, consider the symmetry transformations generated by $P_{\pm}$ and $B_{\pm}$ which represent the newly introduced charges w.r.t. the previous works \cite{Geiller:2020xze, Achour:2021dtj}. They can again be compactly written as
\begin{align}
\tau \rightarrow \tilde{\tau} & = \tau \\
V_1 \rightarrow \tilde{V}_1 (\tilde{\tau}) & = V_1(\tau) + 2 \chi(\tau) \sqrt{V_1}\\
V_2 \rightarrow \tilde{V}_2 (\tilde{\tau}) & = V_2(\tau) + \xi(\tau) \sqrt{V_1}(\tau)  + \chi(\tau) \frac{V_2(\tau)}{\sqrt{V_1}(\tau)}
\end{align}
where $(\xi, \chi)$ are two functions of the $\tau$-coordinate.
The infinitesimal transformation of the action is given by
\be
\delta_{\xi, \chi} S = 2 \epsilon  \ell_p c\int \rd \tau \left\{  \ddot{\xi} \sqrt{V_1} + \ddot{\chi} \frac{V_2}{\sqrt{V_1}}-  \frac{\rd }{\rd \tau} \left( \dot{\xi} \sqrt{V_1} - \dot{\chi} \frac{V_2}{\sqrt{V_1}} - \frac{\epsilon}{2\ell^2_s} (f-\tau)\right) + \O(\chi^2, \xi^2)\right\}
\ee
such that we have a symmetry provided $\ddot{\xi} = \ddot{\chi} =0$. We obtain thus
\be
\chi(\tau) = \alpha_{+} + \alpha_{-} \tau \;, \qquad \xi(\tau) = \beta_{+} + \beta_{-} \tau
\ee
which provides a four-parameter Noether symmetry of the action. The $P$'s and $B$'s thus act as translations and boosts in field space. This concludes the presentation of the symmetry transformations.

\subsubsection{Integrating the dynamics}

We now want to show that the Schr\"{o}dinger charges fully encode the Schwarzschild geometry, i.e. that their knowledge is enough to reconstruct the geometry. It implies that the constants of integration $(C, \tau_1, \tau_0)$ introduced in (\ref{sol}) can be expressed in terms of the conserved charges. To see this, remember that the gauge-fixed metric is given by
\be
\rd s^2 = \epsilon \left( - \frac{\rd \tau^2}{A^2(\tau)} + A^2(\tau) \rd y^2\right) + \ell^2_s B^2(\tau) \rd \Omega^2
\ee
where the $A$ and $B$-fields are given in terms of the dynamical fields $V_1$ and $V_2$ by (\ref{relAB}).
In order to algebraically determine their profiles, one can use the translations and boosts (\ref{bb}) and (\ref{pp}). One finds
\be
\sqrt{V_1} = \frac{1}{4\epsilon c \ell_P} \left( B^{\epsilon}_{-} - \tau P_{-}\right)  \;, \qquad \frac{V_2}{\sqrt{V_1}} = \frac{1}{2\epsilon c \ell_P} \left( B^{\epsilon}_{+} - \tau P_{+}\right) 
\ee
These solutions allows one to reconstruct the profiles of the $A$ and $B$-fields such that
\be
\frac{A^2(\tau)}{4} =  \frac{B^{\epsilon}_{+} - \tau P_{+}}{B^{\epsilon}_{-} - \tau P_{-}}  \;, \qquad 4 \epsilon  \ell_p cB(\tau) = B^{\epsilon}_{-} - \tau P_{-}
\ee
With this profile at hand, one can rewrite the constants of integration in (\ref{sol}) in terms of the conserved charges:
\begin{align}
\frac{\ell^2_s}{c\ell_P} = \frac{1}{Q_{+}} \;, \qquad C = - \frac{P_{-}}{4\epsilon c \ell_P} \;, \qquad \tau_1 = \frac{B^{\epsilon}_{+}}{P_{+}} \;, \qquad \tau_0 =  \frac{B^{\epsilon}_{-}}{P_{-}}
\end{align} 
This provides the dictionary between the observables of the Schr\"odinger algebra and the constants of integration directly entering the expressions of the classical trajectories. This underlines the role of constants of integration as constants of motion, and vice-versa the interpretation of conserved charges as initial conditions for the trajectories.
This concludes the reconstruction of the Schwarzschild geometry from the Schr\"{o}dinger observables. We now present an additional set of symmetries for these two gravitational mini-superspaces.

\section{Superspace conformal isometries and Witt charges}

\label{infsym}
In this section, we show that the two geometrization schemes based on the superspace and the ED lift methods are complementary. The key point is that while Killing symmetries of the superspace are also Killing symmetries of the ED lift, this is no longer true for the conformal Killing isometries. Therefore, each method cannot exhaust all the symmetries present in the system and they have to be used hand in hand (as well as both extended to possibly higher rank killing tensors). To illustrate that point, we now present the infinite dimensional symmetry of the cosmological and black hole mini-superspaces which descends from the infinite conformal isometries of their 2d superspace metrics.

Consider a mechanical system with two degrees of freedom. One can always find a suitable field redefinition such that the action reads
\begin{align}
S[\chi^a ; \tau] =  \int \rd \tau \; g_{ab} (\chi) \dot{\chi}^a \dot{\chi}^b = \int \rd \tau \;  \dot{\chi}_{+} \dot{\chi}_{-}
\end{align}
In terms of these dynamical fields, the field space metric becomes
\be
\rd s^2 = 2\rd \chi_{+} \rd \chi_{-}
\ee
The conformal killing equation leads to the three conditions
\begin{align}
\L_{\xi} g_{ab} = \varphi g_{ab} \;, \qquad  \Rightarrow \qquad 
\left|\begin{array}{l}
\partial_{+} \xi^{-}= 0 \;,\\
\partial_{-} \xi^{+}= 0 \;,\\
\partial_{-} \xi^{-} + \partial_{+} \xi^{+}= \varphi \;,
\end{array}\right.
\end{align}
Therefore, one obtains two families of CKVs given by
\be
X_F^a \partial_a = F(\chi_{+}) \partial_{+} \;, \qquad Z_G^a \partial_a =  G(\chi_{-}) \partial_{-}
\ee
which form two copies of a $\cw$ algebra:
\be
[X_{F}, X_{G}]=2X_{[F,G]} \;,  \qquad [Z_{F}, Z_{G}]=2Z_{[F,G]} \;,  \qquad [X_{F},Z_{G}]=0\,,\qquad
\ee
where $[F,G] = - [F,G] =F G' - GF'$. See \cite{DiFrancesco:1997nk} for more details on conformal invariance in two dimension and the Witt algebra. 
Introducing the conjugated momenta $p_{\pm}$ of the dynamical fields $\chi_{\pm}$, one obtains an infinite tower of conserved charges given by
\be
X_F = F(\chi_{+}) p_{+} \;, \qquad  Z_G = G(\chi_{-}) p_{-}
\ee
The hamiltonian of the system being given by $h = p_{+} p_{-}$ and if it descends from a relativistic action, i.e. if it vanishes on-shell $h \heq 0$, one can check that these sets of $\cw$ charges are indeed weak Dirac observables of the system
\be
\{ X_F,h \} = F' h \heq 0\;, \qquad \{ Z_G, h\} = G' h \heq 0
\ee
Now let us present the explicit construction of the Witt charges for the flat FRW cosmology and the Schwarzschild mechanics.

\subsection{Witt charges for FRW cosmology}

\label{WittFRW}


Consider the charges generating the translations, i.e $P_{\pm}$. Since these charges do not depend on time, they stand as strong Dirac observables of the system. Interestingly, one can construct from them an infinite tower of weak Dirac observables as follows. Consider for example the phase space function
\be
\label{witt}
W^{\pm}_F = \frac{1}{2} \left(\frac{\pi}{\ell_P} \pm \frac{zp}{2} \right) F_{\pm}(\ell_P \phi \pm 2 \log{z})
\ee
where $F_{\pm}$ are arbitrary functions. It is straightforward to show that they represent weak Dirac observables of the system:
\be
\{ W^{\pm}_F , h\} = \left( F_{\pm}' \pm \frac{F_{\pm}}{2}\right) h \; \heq \; 0
\ee
This shows that the  FRW cosmological system is equipped with an infinite set of such observables. They satisfy the following $\cw$ algebras
\begin{align}
\label{algwitt}
& \{ W^{\pm}_F, W^{\pm}_G\}  = W^{\pm}_{[F,G]}  \;, \qquad \text{where} \qquad [F,G] = F' G - G'F \,,\\
& \{ W^{+}_F, W^{-}_G\} =0\,.
\end{align}
Here the parameter functions $F(\tau)$ and $G(\tau)$ are to be understood as vectors fields in $\tau$, then the commutator $ [F,G]$ is simply the Lie derivative.


\medskip

Finally, let us point that demanding strong Dirac observables from  this construction imposes that $F_{\pm}' = \mp F/2$.  In terms of the canonical variables $(\phi, z, \pi, p)$, the resulting charges read
\be
W^{\pm} = \frac{1}{2} \left( \frac{\pi}{\ell_P z} \pm \frac{p}{2}\right) e^{\mp \ell_P \phi/2}
\ee
%
which reproduce the generators of  translations $P_{\pm}$ introduced earlier in (\ref{pb}). 
Then the weak observables for arbitrary parameter functions $F$ and $G$ are interpreted as the equivalent of super-translations, if we compare our construction to the BMS symmetry structure of the dynamics of  asymptotically flat space-times.

\subsection{Witt charges for Schwarzschild mechanics}

\label{wittsch}

Finally, let us construct the infinite tower of $\cw$ charges for  the Schwarzschild mechanics. The super-metric of the Schwarzschild mechanics is given by
\be
\label{supSch}
\rd s^2 = g_{ab} \rd x^a \rd x^b =  \frac{V_2}{V^2_1} \rd V^2_1 - \frac{2\rd V_1 \rd V_2}{V_1} 
\ee
Notice that the Schwarzschild dynamics does not correspond to a null geodesic on this geometry because the action contains a constant potential term. Therefore, depending on the sign of this contribution, the dynamics is either mapped to time-like or space-like geodesics of the super-metric (\ref{supSch}). Thus, as explained in the first section, the CKVs of this geometry do not provide directly conserved charges for the system, and one needs to add corrective terms proportional to the mass of the system, i.e. proportional to the constant potential in the present case.  With this subtlety in mind, we now present the KVs and the CKVs of this super-metric. First, one can check that this field space has constant curvature  such that it admits three Killing vectors. They are given by
\begin{align}
P^{a}_{+} \partial_a = 2\sqrt{V_1} \partial_{V_1} + \frac{V_2}{\sqrt{V_1}} \partial_{V_2} \;, \qquad P^{a}_{-} \partial_a = \sqrt{V_1}\partial_{V_2}\;, \qquad J^{a} \partial_a = V_1 \partial_{V_1}
\end{align}
which correspond to the KVs of the ED lift. Being Killing vectors, they are directly associated to strong Dirac observables of the system and no corrective terms need to be added.

To derive the $\cw$ charges corresponding to the CKV, let us introduce the new coordinates
\be
x_{+} = \sqrt{V_1} \;, \qquad x_{-} = - \frac{V_2}{\sqrt{V_1}}
\ee
such that the (rescaled) super-metric reads
\be
\rd s^2 = \rd x_{+} \rd x_{-}
\ee
The general expressions for the CKVs are given by
\be
X^a_{F} \partial_a = F(x_{+}) \partial_{+} \;, \qquad X^a_G \partial_a = G(x_{-}) \partial_{-}
\ee
where $F$ and $G$ are arbitrary functions. In order to compute the associated charges, we first write the momenta $p_{\pm}$ which are given by
\begin{align}
p_{+} = 2\sqrt{V_1}P_1 + \frac{V_2 P_2}{\sqrt{V_1}} \;, \qquad p_{-} =  -\sqrt{V_1} P_2
\end{align}
They are nothing else but the translational charges entering in the Schr\"{o}dinger algebra. Then, one obtains the following set of observables
\be
X^{+}_F = p_{+} F(\sqrt{V_1}) \;, \qquad X^{-}_G = p_{-} G\left( \frac{V_2}{\sqrt{V_1}}\right)
\ee
As expected, these phase space functions are not conserved on shell. A straightforward computation shows that
\be
\{ X^{+}_F ,  h\} = F' \tilde{h} \;\heq  - \frac{c\ell_P F'}{\ell^2_s} \;, \qquad \{ X^{-}_G, h\} = 2G' \tilde{h} \;\heq  - \frac{2c\ell_P G'}{\ell^2_s}
\ee
where $\tilde{h}$ is the shifted hamiltonian whose value can be interpreted as the effective mass of the test particle moving on the field space. To obtain well defined weak Dirac observables, we slightly correct the charges as follows. Consider
\begin{align}
W^{+}_F & = X_F - \frac{\alpha}{2\sqrt{V_1}P_2} F(\sqrt{V_1}) \;, \;\;\qquad \text{with} \qquad 
\alpha = 2\epsilon c^2 \frac{\ell^2_P}{\ell^2_s} \;,\\
W^{-}_G & = X_G \left[ 2 -  \frac{\beta}{\tilde{h}} \right] G\left( \frac{V_2}{\sqrt{V_1}}\right), \qquad \text{with} \qquad \beta = - \frac{c\ell_P}{\ell^2_s}
\end{align}
With these corrections, the modified charges now satisfy
\be
\{ W^{+}_F ,  h\} = F' \left( \tilde{h} + \frac{c\ell_P}{\ell^2_s} \right) \heq \; 0\;, \qquad \{ W^{-}_G, h\} = 2G' \left( \tilde{h} + \frac{c\ell_P}{\ell^2_s} \right) \heq \; 0
\ee
Now, using that $P_{\pm}$ are the two translation charges of the Schr\"{o}dinger algebra introduced in (\ref{bb}) and (\ref{pp}) which are related to the hamiltonian via $P_{+} P_{-} = - 4\epsilon  \ell_p cQ_{+}$, we can rewrite the Witt charges as follows
\begin{align}
W^{+}_{F} & =  F (\sqrt{V_1}) \left[ P_{+} - \frac{\alpha}{P_{-}} \right]  \;,  \\
W^{-}_G & = G\left(\frac{V_2}{\sqrt{V_1}}\right)  \left[ P_{-} - \frac{\beta P_{-}}{ Q_{+}} \right]   \;, 
\end{align}
Each set of charges form a $\cw$ algebra, i.e.
\begin{align}
\{ W^{\pm}_F, W^{\pm}_G\} & = W^{\pm}_{[F,G]} 
\end{align}
but they  do not commute with each other. Indeed, one can show that
\begin{align}
\label{noncom}
\{ W^{+}_F, W^{-}_G\} & = \alpha \beta  \frac{4 \epsilon c \ell_P}{P_{+} P_{-}} \left( \frac{G'F}{P_{-}} + \frac{G F'}{P_{+}} \right) + \alpha \frac{G'F}{2P_{-}} - 4 \epsilon  \ell_p c\beta \frac{G F'}{P_{+}}\quad\ne0\,.
\end{align} 
We observe that the anomalies originate from the corrective terms, labelled by $(\alpha, \beta)$, that we have introduced in order to obtain well defined conserved charges.
From that perspective, these $\cw$ charges are different from the ones derived for the flat  FRW case. This can be understood from the potential contribution of the symmetry-reduced action of each model. In the  FRW case, the action does not contain any potential term, such that the cosmological dynamics can be mapped to null geodesics in field space. It follows that any CKVs of the cosmological super-metric
 is a well defined charge of the dynamics. On the contrary, the symmetry reduced action (\ref{acSch}) for the black hole mechanics contains a constant potential. At the level of the field space, this term plays the role of a non-vanishing mass and the black hole dynamics is mapped onto space-like or time-like geodesics on the super-metric. As explained in the first section, for such model, the CKVs of the super-metric do not provide  anymore conserved charges of the system. Instead, one needs to add corrective terms proportional to the "mass" of the system to recover well defined weak observables from the CKVs.  As we have seen, a consequence of adding these corrective terms is that, in general, the two copies of $\cw$ charges one can derive do not commute anymore. See \cite{Dimakis:2022pks} for a related discussion.



\section{Discussion}
\label{disc}

\bigskip

We have presented new dynamical symmetries of cosmological and black hole mini-superspaces. We have considered two models: i) the  FRW cosmology filled with a massless scalar field, and ii) the Schwarzschild black hole mechanics. 
In order to systematically investigate the symmetries of these reduced gravitational models, we have employed the geometrization approach, which allows one to recast the (gauge fixed) dynamics of these gravitational systems as a geodesic on an auxiliary background, i.e. a "second geometrization" of (homogeneous) gravity. We have considered two such geometrization schemes, based on the well-known field space metric on the one hand, and on its extended version known as the Eisenhart-Duval (ED) lift on the other hand.  

Using this strategy, we have shown that these two gravitational models are invariant under the two-dimensional Schr\"{o}dinger group,  well known as the key symmetry in classical and quantum mechanics. Nevertheless, a key difference with the standard Schr\"{o}dinger symmetry shows up in the existence of boost generator (\ref{pb}) which replaces the standard compact rotation generator such that the algebra of observables decomposed into
\be
\sh(2) =( \sl(2,\mathbb{R})\times \so(1,1) )  \ltimes( \mathbb{R}^2 \times \mathbb{R}^2)
\ee
 While the existence of this symmetry can be expected since our gravitational systems can be recast as free particles on an auxiliary conformally flat background, the explicit expression of their Schr\"{o}dinger charges requires the use of the ED lift method. The identification of this charge algebra through this systematic procedure has several interesting outcomes. First, it provides a geometrical origin for the conformal symmetries initially identified in \cite{Pioline:2002qz, BenAchour:2019ufa, BenAchour:2020njq, Achour:2021lqq} which correspond to the SL$(2,\mathbb{R})$ sector of Schr\"{o}dinger group. Second, it gives an extended charge algebra to guide the quantization of these gravitational mini-superspaces. In particular, the Casimirs which vanish at the classical level are expected to encode the quantum corrections due to the higher moment of the wave function of the geometry in the quantum theory encoding the quantum fluctuations of geometric observables. Thus they provide an algebraic characterization of the deviation from the classical to the quantum geometry. 

The realization of this symmetry in reduced gravitational models might seem puzzling at first sight. Indeed, the Schr\"{o}dinger group is a well-known feature of non-relativistic massive systems \cite{Nied, Horvathy:2009kz, Duval:2009vt, Duval:2012qr}. Its realization in a relativistic gravitational system, being one-dimensional, is rather non-trivial. However, as shown in Section~\ref{sym}, the Schr\"{o}dinger symmetry is a field-space symmetry in our context. One way to make this difference manifest is by comparing the quantities which transform under the scaling symmetry. In classical mechanics, where the Schr\"{o}dinger symmetry is interpreted as a space-time symmetry, time and space are rescaled as
\be
t \rightarrow \lambda t \;, \qquad x^i \rightarrow \lambda^2 x^i
\ee
In the present context, the spatial coordinates $x^i$ is replaced by specific dynamical fields $\chi^i$ entering in our homogeneous geometry. In FRW cosmology, this role is played by the (appropriate power of the) scale factor, while for the black hole, it corresponds to a combination of the metric components. Therefore, at the level of the space-time metric, the conformal sector of the Schr\"{o}dinger symmetry translates into an anisotropic Weyl rescaling of the metric components. Moreover, the relativistic nature of our gravitational system can be linked with the fact that the generator $J$ in (\ref{pb}) is not a rotation but a boost (in field space). Indeed, while our gravitational system are recast into a mechanical system, i.e. a particle moving on a 2d auxiliary space, the relativistic nature of the system manifests through the lorentzian signature of the field space. This is ultimately the key difference with a standard mechanical system and why we inherit a boost instead of rotation generator in the Schr\"{o}dinger algebra of observables we have identified.

We have further shown that using the standard field space approach reveals additional symmetries. In particular,  the field space of the considered models are two-dimensional, and thus possess an infinite number of independent conformal Killing vectors fields. This translates into an infinite set of conserved charges for these two systems which can be organized in two copies of a $\cw$ algebra. If the gravitational system can be mapped to a free particle, the two copies will commute as it is the case for FRW cosmology. But in the case of the Schwarzschild mechanics, where the system is mapped onto a particle with a constant potential, deformations of the charges prevent the two copies to commute with each other, paralleling the results presented in \cite{Dimakis:2022pks}. Whether these symmetries can be useful to the quantization of these systems remain to be shown. Yet, quantizing these two gravitational models amounts to capturing the dynamics of their wave function on a 2d Lorentzian manifold which is by construction conformally invariant. This raises the question whether  the techniques from 2d lorentzian CFT can be used to tackle the quantization of these symmetry-reduced gravitational systems and discuss the fate of the singularity in this context from a new perspective.

Finally, an interesting perspective suggested by this work is to construct non-linear extensions of the standard Wheeler-de Witt (WdW) quantization based on symmetries. Indeed, the Schr\"{o}dinger symmetry is well known to be preserved for suitable non-linear extensions of the free Schr\"{o}dinger equation. This is typically the case for the 2d Gross-Pitaevskii equation or the 1d Tonks-Gerardeau gas which are both used to describe Bose-Einstein condensates \cite{Ghosh:2001an}. The present symmetry therefore suggests that one can extended the standard (linear) WdW equation by a non-linear interaction whose form is protected by the Schr\"{o}dinger symmetry identified in our work. This non-linear extension will play the role of a symmetry-protected quantum correction for our gravitational systems. Understanding its effects on the dynamics of the wave function of the universe or for quantum black hole models is an interesting goal we wish to develop in the future. The present work provides a preliminary study for that future project. Let us point that non-linear extensions of quantum cosmology equations, thus hydrodynamics-like equations for a wavefunction on field space, have been in fact obtained from an hydrodynamic approximation of the fundamental quantum gravity dynamics in the context of TGFT condensate cosmology \cite{Gielen:2013naa, Gielen:2016dss, Pithis:2019tvp}, aiming at the extraction of effective cosmological dynamics from full quantum gravity. They have also been argued to be a general outcome of a coarse-graining procedure applied to the dynamics of quantum gravity degrees of freedom. Our results on field symmetries in a cosmological context can be seen as supporting this perspective. More important, they can provide rigorous symmetry-based guidelines for developing further this research direction. This approach might also open the road to construct new dictionaries between quantum gravitational models and non-linear Schr\"{o}dinger systems which could reveal useful for developing analogue quantum gravity simulations in condensed matter systems.

\bigskip
\bigskip
 \textbf{Acknowledgments}\smallskip

 \medskip
The work of J. Ben Achour is supported by the Alexander von Humboldt foundation \\AOSt.820 022-6. D. Oriti acknowledges funding from DFG research grants OR432/3-1 and \\ OR432/4-1.
\newpage 
\appendix


\section{Einsenhart-Duval lift of  FRW cosmology}
\label{app:A}
\label{so42}

In this appendix, we explicitly show that the conformal isometries of the Eisenhart-Duval lift associated to the flat  FRW model form indeed an $\so(4,2)$ Lie algebra.
Consider the flat metric and the coordinates
\begin{equation}
\eta_{\alpha\beta}=\begin{pmatrix}
-1 & 0 & 0 & 0\\
0 & 1 & 0 & 0\\
0 & 0 & 0 & 1\\
0 & 0 & 1 & 0
\end{pmatrix}
\qquad\qquad\qquad
x^\alpha=\begin{pmatrix}
-u+w \\
u+w\\
ze^{+\phi}\\
ze^{-\phi}
\end{pmatrix}
\end{equation}
A given vector field can be decomposed as
\be
\begin{array}{c}
\partial_\alpha=\frac{\partial}{\partial x^\alpha}=\begin{pmatrix}
-\partial_u +\partial_w ,& \partial_u +\partial_w ,& e^{-\phi}\left[\partial_z+\frac{\partial_\phi}{z}\right] ,& e^{+\phi}\left[\partial_z-\frac{\partial_\phi}{z}\right]
\end{pmatrix}
\end{array}
\ee
The generators of the $\so(4,2)$ Lie algebra are given by
\begin{align}
\text{4 translations}& & &T_\alpha=\partial_\alpha \label{translations}\\
\text{6 rotations}& & &M_{\alpha\beta}=x_\alpha\partial_\beta -x_\beta\partial_\alpha\\
\text{1 dilation}& & &D=x^\alpha \partial_\alpha\\
\text{4 inversions}& & &K_\alpha=2x_\alpha x^\beta\partial_\beta - x^\beta x_\beta \partial_\alpha \label{inversion}
\end{align}
which satisfy the standard commutation relations
\begin{equation}\begin{split}
&\left[M_{\alpha\beta} M_{\gamma\delta}\right]=\eta_{\alpha\delta}M_{\beta\gamma}+\eta_{\beta\gamma}M_{\alpha\delta}+\eta_{\alpha\gamma}M_{\delta\beta}+\eta_{\beta\delta}M_{\gamma\alpha}\\
&\left[T_\alpha, T_\beta\right]=0; \qquad \left[T_\alpha, M_{\beta\gamma}\right]=\eta_{\beta\alpha}T_\gamma-\eta_{\gamma\alpha}T_\beta\\
&\left[K_\alpha, K_\beta\right]=0; \qquad \left[K_\alpha, M_{\beta\gamma}\right]=\eta_{\beta\alpha}K_\gamma-\eta_{\gamma\alpha}K_\beta\\
&\left[D, K_\alpha\right]=K_\alpha; \qquad \left[D, M_{\alpha\beta}\right]=0; \qquad \left[T_\alpha, D\right]=T_\alpha\\
&\left[T_\alpha, K_\beta\right]=2\left(\eta_{\alpha\beta}D -M_{\alpha\beta}\right)
\end{split}
\end{equation}
In term of the coordinates of the lift, one can write the rotational generators $M_{\alpha\beta}$ as follows
\begin{align}
M_{01}&=2\left(u\partial_u -w\partial_w\right) \\
M_{0\pm}&=e^{\mp\phi}\left[\left(u-w\right)\left(\partial_z \pm\frac{\partial_\phi}{z}\right)-z\left(-\partial_u+\partial_w\right)\right] \\
M_{1\pm}&=e^{\mp\phi}\left[\left(u+w\right)\left(\partial_z \pm \frac{\partial_\phi}{z}\right)-z\left(\partial_u +\partial_w\right)\right] \\
M_{+-}&=-2\partial_\phi
\end{align}
while the generators $K_{\alpha}$ are given by
\begin{align}
K_0&=4\left(u^2 +\frac{z^2}{2}\right)\partial_u -4\left(w^2+\frac{z^2}{2}\right)\partial_w +4z\left(u-w\right)\partial_z \\
K_1&=4\left(u^2 -\frac{z^2}{2}\right)\partial_u +4\left(w^2-\frac{z^2}{2}\right)\partial_w +4z\left(u+w\right)\partial_z \\
K_\pm&= e^{\mp\phi}\left[4z\left(u\partial_u +w\partial_w\right)+\left(2z^2 -4uw\right)\partial_z \mp\frac{\left(2z^2+4uw\right)}{z}\partial_\phi\right]
\end{align}
Finally, the dilatation takes the form
\begin{align}
D=2u\partial_u +2w \partial_w +2z\partial_z
\end{align}
At this point, we can reorganize the $\so(4,2)$ generators to extract the Schr\"{o}dinger ones. These vector fields are given by
\begin{equation*}
Q_+=\frac{1}{8} \left(K_1+K_0\right)\qquad\quad
Q_0=-\frac{1}{4}\left(D+M_{01}\right)\qquad\quad
Q_-= \frac{1}{2}\left(T_1 - T_0\right)
\end{equation*}
together with 
\begin{equation*}
P_\pm=P_\pm\qquad \qquad B_\pm=-\frac{1}{2}\left(M_{1\pm}+M_{0\pm}\right) \qquad\quad
J=-\frac{1}{2}M_{+-}
\end{equation*}
Finally, the dual generators coming from the switch $u \leftrightarrow w$ in the lifted metric read
\begin{equation*}
\tilde{Q}_+=\frac{1}{8}\left(K_1-K_0\right)\qquad\quad
\tilde{Q}_-=\frac{1}{2}\left(T_1+T_0\right) \qquad \qquad \tilde{B}_\pm=-\frac{1}{2}\left(M_{1\pm}-M_{0\pm}\right)
\end{equation*}
The $\so(4,2)$ algebra is completed by the three remaining generators
\begin{equation*}
\mathcal{O}=-\frac{1}{2}M_{01}\qquad\quad
Y_\pm=\frac{1}{2}K_\pm
\end{equation*}
This concludes the presentation of the $\so(4,2)$ algebra of conformal isometries of the Eisenhart-Duval lift associated to the flat  FRW cosmology filled with a massless scalar field.

\section{Eisenhart-Duval lift of Schwarzschild mechanics}

\label{app:C}

In this appendix, we provide the explicit expressions of the CKVs of the Eisenhart-Duval lift associated to the black hole as well as the CKVs of the Schwarzschild field space.
The  gauge-fixed action for the Schwarzschild mechanics is given by (\ref{gaugefixSch}). The potential term being a mere constant, one can omit it and consider the modified mechanical action 
\be
S_{\epsilon} [V_1, V_2; \tau]= \epsilon  \ell_p c\int \rd \tau \left[  \frac{V_2 \dot{V}^2_1- 2 V_1 \dot{V}_1 \dot{V}_2}{2V^2_1}  \right]
\ee
Then, the ED lift is a four dimensional manifold with line element
\be
\rd s^2 = g_{AB} \rd X^A \rd X^B = 2  \rd u \rd w +\epsilon c  \ell_P \left(  \frac{V_2}{V^2_1} \rd V^2_1 - \frac{2\rd V_1 \rd V_2}{V_1} \right)
\ee
Just as for the  FRW cosmological system, this manifold turns out to be conformally flat, i.e its Weyl and Cotton tensors vanish, and it is thus maximally symmetric, i.e. it possesses fifteen CKVs\footnote{Notice that if we had kept the constant term in the action, the ED lift would have been given by 
\be
\rd s^2 = \frac{2\epsilon}{\ell^2_s} \rd u^2 +   2\rd u \rd w +\epsilon  \ell_p c\left(  \frac{V_2}{V^2_1} \rd V^2_1 - \frac{2\rd V_1 \rd V_2}{V_1} \right)
\ee
One can see that the potential term being constant, $\partial_u$ is still a Killing vector. Moreover, one can show that this additional term does not spoil the conformally flatness of the lift such that this alternative version admits again fifteen CKVs. While the form of these CKVs is slightly different starting from this lift, the conserved charges one builds from the CKVs are the same.}. They can be organized w.r.t. the commutation properties with the constant null vector field $N^A\partial_A = \partial_w$.

The CKVs which commute with $\partial_w$ are given by 
\begin{align}
         Q_0^{A}\partial_{A}  & = V_2 \partial_{V_2} + u \partial_u \,,  \\
         Q_{-}^{A} \partial_{A} & = 2 V_2 \partial_w + \frac{2u}{\epsilon c} \left( V_1 \partial_{V_1} + V_2 \partial_{V_2} \right) + \frac{1}{\epsilon c} u^2 \partial_u  \,, 
\end{align}
and act on the metric of the lift as
\be
\L_{Q_{0}} g_{\mu\nu} = g_{\mu\nu}\;, \qquad \L_{Q_{-}} g_{\mu\nu} =\frac{4u}{c} g_{\mu\nu} 
\ee
We also identify six Killing vectors which are given by
\begin{align}
\begin{array}{c}
 \left|
    \begin{array}{l}
         P_{+}^{A} \partial_{A}  = 
          2\sqrt{V_1} \partial_{V_1} + \frac{V_2}{\sqrt{V_1}} \partial_{V_2}
            \,, \\
         P_{-}^{A} \partial_A = 
         \sqrt{V_1} \partial_{V_2}  \,,  \\ 
    \end{array}
\right. 
\qquad
 \left|
    \begin{array}{l}
         B_{+}^{A} \partial_{A}  = \frac{2V_2}{\sqrt{V_1}} \partial_w + \frac{u}{\epsilon c} \left( 2\sqrt{V_1} \partial_{V_1} + \frac{V_2}{\sqrt{V_1}} \partial_{V_2}\right) \,,\\        
         B_{-}^{A} \partial_{A}  = 2 \sqrt{V_1} \partial_w + \frac{u}{\epsilon c} \sqrt{V_1} \partial_{V_2} \,,  \\
    \end{array}
\right. \\ \\
Q_{+}^{\alpha} \partial_{\alpha}  = \partial_u \,,  \qquad
         J^{\alpha} \partial_{\alpha}  = 
          V_1 \partial_{V_1} \;,
\end{array}
\end{align}
Together with the null vector $N^A\partial_A = \partial_w$, they form the Schr\"{o}dinger algebra from which one can derive the Schr\"{o}dinger observables presented in Section~\ref{schsch}.

Three other CKVs are easily identified as
\begin{align}
         \tilde{Q}_{-}^{A} \partial_{A} & = 2 V_2 \partial_u + \frac{2w}{\epsilon c} \left( V_1 \partial_{V_1} + V_2 \partial_{V_2} \right) + \frac{1}{\epsilon c} w^2 \partial_u  \,, \\
         \tilde{B}_{+}^{A} \partial_{A}  &= \frac{2V_2}{\sqrt{V_1}} \partial_u + \frac{w}{\epsilon c} \left( 2\sqrt{V_1} \partial_{V_1} + \frac{V_2}{\sqrt{V_1}} \partial_{V_2}\right)\\
         \tilde{B}_{-}^{A} \partial_{A}  & = 2 \sqrt{V_1} \partial_u + \frac{w}{\epsilon c} \sqrt{V_1} \partial_{V_2} .
\end{align}
The first one acts on the lift as
\be
\L_{\tilde{Q}_{-}} g_{\mu\nu} =\frac{4w}{c} g_{\mu\nu} \;, 
\ee
They are directly obtained as the dual of the CKVs $Q_{-}^A \partial_{A}$ and $B^A_{\pm} \partial_A$ under the switch $u \leftrightarrow w$. The dual of the CKV $Q_{0}^A \partial_{A}$ is recovered thanks to the generator of dilatation in the $(u, v)$-plane $\mathcal{O}=u\partial_u - w\partial_w$. Finally, the last two CKVs are
\begin{align}
Y^A_{+} \partial_A &= \frac{2V_2}{\sqrt{V_1}}\left(u\partial_u+w\partial_w\right) +2uw\sqrt{V_1}\partial_{V_1}+\left(uw+2V_2\right)\frac{V_2}{\sqrt{V_1}}\partial_{V_2}\\
Y^A_{-} \partial_A &= 2\sqrt{V_1}\left(u\partial_u +w\partial_w\right) +4V_1\sqrt{V_1}\partial_{V_1}+\left(uw+2V_2\right)\sqrt{V_1}\partial_{V_2}
\end{align}
Altogether, these fifteen vectors fields from the $\so(4,2)$ Lie algebra associated to the Schwarzschild black hole mechanics.

\end{document}